\documentstyle[12pt]{article}

\begin{document}

\baselineskip 18pt

\newcommand{\sheptitle}
{Dynamical Electroweak Symmetry Breaking
\footnote{This revised version appears in
Reports on Progress in
Physics {\bf 58} (1995) 263.}}

\newcommand{\shepauthor}
{Stephen F King\footnote{e-mail address king@soton.ac.uk}}

\newcommand{\shepaddress}
{Physics Department, University of Southampton\\Southampton, SO9 5NH, U.K.}

\newcommand{\shepabstract}
{We review the status of and recent developments in
dynamical electroweak symmetry breaking,
concentrating on the ideas of technicolour
and top quark condensates.
The emphasis is on the essential physical ideas and
experimental implications rather than on detailed
mathematical formalism.
After a general overview of the subject,
we give a first introduction to technicolour,
and extended technicolour, illustrating the ideas
with a simple (unrealistic) model.
Then we review the progress that has been made with
enhancing the technicolour condensate, using the Schwinger-Dyson
gap equation. The discussion includes the so-called walking technicolour
and strong extended technicolour approaches.
We then turn to the experimental prospects of
technicolour models, including longitudinal gauge boson
scattering experiments at the LHC, the detection of
pseudo-Goldstone bosons and the hints about electroweak symmetry
breaking which comes from precision measurements at LEP.
We also discuss a low-scale technicolour model,
which has experimental signatures at LEP and the Tevatron.
Finally we turn to the idea of the top quark condensate.
After reviewing the basic ideas of this approach,
we turn to some extensions of these ideas involving the idea
of fourth family condensates, and the role of irrelevant
operators. }

\newcommand{\beq}{\begin{equation}}
\newcommand{\eeq}{\end{equation}}
\newcommand{\beqarr}{\begin{eqnarray}}
\newcommand{\eeqarr}{\end{eqnarray}}
\newcommand{\psibarpsi}{\mbox{$<\bar{T}T>_{M_{ETC}}$}}
\newcommand{\alphaetc}{\mbox{$\alpha_{ETC}$}}

\begin{titlepage}
\begin{flushright}
SHEP 93/94-27 \\
hep-ph/9406401
\end{flushright}
\vspace{.4in}
\begin{center}
{\large{\bf \sheptitle}}
\bigskip \\ \shepauthor \\ \mbox{} \\ {\it \shepaddress} \\ \vspace{.5in}
{\bf Abstract} \bigskip \end{center} \setcounter{page}{0}
\shepabstract
\end{titlepage}

\newpage

{\bf \Large Contents}

\vspace{0.25in}

{\bf 1. Introduction}

   1.1 What is dynamical symmetry breaking?

   1.2 The standard model

   1.3 Overview of dynamical electroweak symmetry breaking

\vspace{0.25in}

{\bf 2. Technicolour}

   2.1 The minimal technicolour model

   2.2 Extended technicolour

   2.3 A simple extended technicolour model

\vspace{0.25in}

{\bf 3. Condensate Enhancement}

   3.1 Schwinger-Dyson gap equation

   3.2 Walking technicolour

   3.3 Strong extended technicolour

\vspace{0.25in}

{\bf 4. Experimental Prospects}

   4.1 Longitudinal W and Z scattering at the LHC

   4.2 Pseudo-Goldstone bosons from a single techni-family

   4.3 Precision Electroweak Measurements

   4.4 Low scale technicolour

\vspace{0.25in}

{\bf 5. Top Quark Condensates}

   5.1 Four-fermion theory

   5.2 Renormalisation group approach

   5.3 A fourth family after all?

   5.4 Irrelevant operators

\vspace{0.25in}

{\bf 6. Conclusion}

\newpage

\section{Introduction}

\subsection{What is dynamical symmetry breaking?}

The subject of dynamical symmetry breaking has applications to
many branches of physics (e.g. superconductivity, superfluidity,
nucleon pairing, chiral dynamics of hadrons). This review article
will not be concerned with any of these applications but instead
will be concerned with the application of dynamical symmetry
breaking to the problem of understanding the origin of
the {\em electroweak} part of the masses
of elementary particles.\footnote{ The part of the quark masses
generated by their strong interactions will not concern us here.}
However it may be instructive to begin by tracing the development of
dynamical symmetry breaking in an area of physics other than particle
physics, in order to get a feel for what dynamical symmetry breaking
is about. To this end we shall briefly discuss the phenomenon of
superconductivity.

The history of superconductivity began in 1908 at Leiden in Holland
when Kammerlingh Onnes succeeded in liquefying helium.
He not only found that it boiled at 4.2 K, but three years later
discovered that the resistance of a sample of mercury became
negligible quite abruptly at this temperature -- i.e. it became
superconducting. As the supply of liquid helium became
plentiful many metallic alloys were found to exhibit the same
behaviour below their individual critical temperature $T_c$.
The superconducting phase transition was later successfully described
by a theory of Ginzburg and Landau (1950). According to this
theory a macroscopic order parameter,
corresponding essentially to the wavefunction of the superconducting
charges, acquires a nonzero vacuum expectation value (VEV) in the
superconducting state. Although this theory was quite successful
it was later superceded by a theory of Bardeen, Cooper and Schrieffer
(BCS) (1962). In the microscopic BCS theory the dynamical origin
of the order parameter is identified with the formation of
bound states of elementary fermions, namely Cooper pairs of
electrons. Each Cooper pair has zero spin and and zero momentum
and together these pairs form a zero temperature condensate of bosons.
At a finite temperature below $T_c$ some of the pairs are thermally
excited across a small energy gap to form quasi-particle
excitations, as shown in figure 1.
According to BCS theory this energy gap $\Delta$
falls from its maximum value $\Delta (0)$ to zero at $T_c$
in a characteristic way, as shown in figure 2.

As we shall see shortly we have a very successful theory
of particle physics known as the standard model, in which
the symmetry is broken by the VEV of an order parameter
analogous to the order parameter of the Ginzburg-Landau model.
The question which will concern us here
is whether the standard model will one day be superceded by
some dynamical theory analogous to the BCS theory in which
the electroweak symmetry is broken by a condensate of fermion pairs.

\subsection{ The Standard Model}

The standard model of particle physics is based on the
gauge group,

\beq
SU(3)_C \otimes SU(2)_L \otimes U(1)_Y
\eeq

The standard model involves three families of fermions,
the quarks and leptons,
whose left (L) and right (R) handed components
transform under the gauge group as,

\beq
\begin{array}{ccl}
{Q_L}^i= \left( \begin{array}{c}U_L \\ D_L \end{array}
\right)^i &\sim& (3,2,1/6)\\
{U_R}^i &\sim& (3,1,2/3)\\
{D_R}^i &\sim& (3,1,-1/3)\\
{L_L}^i= \left( \begin{array}{c}\nu_{L} \\ E_L \end{array}
\right)^i &\sim& (1,2,-1/2)\\
{E_R}^i &\sim& (1,1,1)
\end{array}
\eeq

where $i=1 \ldots 3$ labels the three families of quarks and leptons,
and in my convention the electric charge generator is given by
\beq
Q=T_{3L}+Y
\eeq
where $T_{3L}$ is the third component of weak isospin
$SU(2)_L$ and $Y$ is the hypercharge generator.

In addition the standard model introduces a complex scalar
doublet,

\beq
H = \frac{1}{\sqrt{2}}
\left( \begin{array}{c}w_1 + iw_2 \\ h_0 + iw_0 \end{array}
\right) \sim (1,2,1/2)
\eeq

where the VEV of the neutral scalar component $<h^0>=v=246$
GeV spontaneously breaks the electroweak symmetry
down to electromagnetism,

\beq
SU(2)_L \otimes U(1)_Y \longrightarrow U(1)_Q
\eeq

The term in the Lagrangian responsible for the $W^{\pm}$ and $Z^0$
masses is the kinetic term for the Higgs doublet,
\beq
(D_{\mu}H)^{\dagger}(D^{\mu}H),
\eeq
where the covariant derivative is defined as,
\beq
D_{\mu}=\partial_{\mu} + ig{W_{\mu}}^aT^a + ig'YB_{\mu}
\eeq
and $T^a$ ($a=1 \ldots 3$) are the $SU(2)_L$ generators.
The combination $(W_1 - iW_2)/\sqrt{2}$ is the $W^+$ field
and receives a mass squared,
\beq
M_W^2=g^2v^2/4
\eeq
while the combination
\beq
Z=\frac{gW_3 -g'B}{\sqrt{g^2+g'^2}}
\eeq
receives a mass squared,
\beq
M_Z^2=(g^2+g'^2)v^2/4
\eeq
and the orthogonal combination is the massless photon field,
\beq
A=\frac{g'W_3 +gB}{\sqrt{g^2+g'^2}}
\eeq
The weak mixing angle is defined by $\tan \theta_W=g'/g$
so that we have the mass relation
\beq
M_W^2/M_Z^2=\cos^2 \theta_W
\eeq
This mass relation has been tested experimentally to
better than 1\% accuracy.

The terms in the Lagrangian responsible for the
fermion masses are the so called Yukawa couplings,
\beq
-g_{ij}\bar{D}_{iR}H^{\dagger}Q_{jL}
-h_{ij}\bar{U}_{iR}\tilde{H}^{\dagger}Q_{jL}
-f_{ij}\bar{E}_{iR}H^{\dagger}L_{jL} +h.c.
\eeq
where $\tilde{H}=i\sigma_2 H^{\star}$,
where $\sigma_2$ is the second Pauli matrix,
is the charge congugate Higgs doublet.
When the Higgs doublet receives its VEV,
fermion masses result.
Thus the Higgs mechanism achieves two things simultaneously:
it provides $W,Z$ masses {\em and} fermion masses.
According to the Higgs mechanism,
the three pseudoscalar degrees of freedom $w_i$ are
eaten by ${W^{\pm}}_L, {Z^0}_L$ and the shifted $h_0$ field
remains in the spectrum as a physical neutral scalar particle
-- the Higgs boson.
More details concerning the Higgs boson can be found in
the review by Cahn (1989).

Experimentally the standard model is in remarkably good shape.
All the fermions of three families of quarks and leptons
have been seen experimentally, including the recent tentative
discovery of the top quark
with a mass of around 174 GeV by Abe {\it et al} (1994).
The standard model is perfectly consistent
with high precision electroweak measurements
which are sensitive to radiative corrections (Ellis {\it et al} 1992).
However the Higgs boson has so far not been detected experimentally,
with LEP putting a lower limit on its mass of about 60 GeV.
This lack of experimental observation of the Higgs boson
is one of the main motivations for considering models
of dynamical electroweak symmetry breaking.

Theoretically the standard model is on less secure ground.
The standard model involves elementary scalar fields,
and these are associated with quadratic divergences at the
one-loop level. Of course such divergences are renormalised away
and the renormalised theory is finite and well behaved.
The problem arises when the standard model is embedded into
some larger structure involving a mass scale $M$ much larger
than the electroweak scale characterised by the VEV $v$.
For example $M$ might be identified with a scale of grand unification
with $M\approx 10^{16}$ GeV. In such a framework the Higgs sector
responsible for breaking the larger gauge symmetry at the scale
$M$ and the Higgs sector responsible for breaking electroweak
symmetry at the scale $v$ cannot be kept distinct,
and communicate through one-loop radiative corrections.
The hierarchy of mass scales can then only be maintained
at the one-loop level by fine-tuning the basic Higgs parameters
of the theory to an accuracy of about 24 decimal places in this example.
Such fine-tuning arises because of the quadratic nature of the scalar
divergences. Furthermore the fine-tuning must be re-done at every order of
perturbation theory. There are two solutions to this
so-called hierarchy problem. The first is to introduce an
$N=1$ supersymmetry which tames the quadratic divergences
into benign logarithmic divergences, due to the
fermionic partners to the scalars. The second, which is the
subject of the present review, is to banish elementary scalars
from the theory, and to replace their effect by fermion dynamics --
the dynamical symmetry breaking approach.

\subsection{Overview of dynamical electroweak symmetry breaking}

Although the standard model provides
a remarkable description of all known experimental data,
there is as yet no direct evidence for the existence of the Higgs
boson. Of course one may argue that the eaten Goldstone bosons (GB's)
of the Higgs mechanism have already been detected, since they may
(in a limit which we shall make precise later) be identified
with the longitudinal (L) polarisation states of the W and Z bosons.
However the key question of electroweak symmetry breaking
is whether these eaten GB's are composite or elementary.
Such questions concerning compositeness must always
be related to some momentum or distance scale, since
systems which appear to be elementary at some scale
may in fact appear to be composite at some higher momentum
(shorter distance) scale. We shall chiefly be interested
in whether the Higgs sector turns out to be composite on
the TeV scale, which will be probed by the next generation of colliders.
We emphasise that even if compositeness effects are revealed
on the TeV scale, this does not imply that the Higgs mechanism
is inoperative. It simply means that the Higgs mechanism is
implemented {\em dynamically}, i.e. that there is dynamical
electroweak symmetry breaking (DEWSB). In other words, there are no
elementary scalars and their effect is replaced by the dynamics
of fermions. Usually this implies some strong interactions between
the fermions in order for a composite fermion-antifermion bound
state to be formed which will play the role of the Higgs scalar fields.

The simplest example of DEWSB is to invent
a new strong confining gauge force called
technicolour (TC) with which to
bind new fermion-antifermion pairs into composite
systems whose behaviour resembles that of the Higgs field.
TC was reviewed by Farhi and Susskind (1981).
Some of the material discussed by Farhi and Susskind
is condensed into sections 2.1, 2.2 and 4.2 of the present article.
The remaining sections of the present article
will focus on more recent developments.
In the TC approach
the new fermions are called technifermions, and the whole
theory is modelled on the known behaviour of quarks
in quantum chromodynamics (QCD) -- but scaled up to the TeV scale.
It turns out that TC by itself is not sufficient to provide
fermion masses. One way forward is to embed the TC gauge group
into a larger gauge group known as extended technicolour (ETC)
(section 2.2).
However we shall see
it is not an easy task to describe the quark and lepton
mass spectrum without running into phenomenological problems.
These problems include the flavour-changing neutral current (FCNC)
problem, problems with massless Goldstone bosons
(or pseudo-Goldstone bosons (PGB's) which are too light).
These problems have thwarted attempts to construct ETC models,
and to date there is no accepted standard ETC model in the
literature. Some recent model building attempts are discussed
in section 2.3.
There are alternatives to ETC which can lead to fermion
masses. For example one may re-introduce the scalar doublet in
order to communicate electroweak symmetry breaking to the
quarks and leptons via Yukawa couplings
(Simmons 1989, Samuel 1990, Kagan and Samuel 1991).
However this appears to be somewhat of a retreat since
the motivation for technicolour is to banish elementary
scalars. However one may envisage a natural scenario involving both
technicolour and supersymmetry (Dimopoulos and Raby 1981).
We shall not discuss this possibility any further here.

Recently ETC has staged a comeback
due to a lot of exciting progress with the above problems.
In section 3 we shall discuss
the recent progress in dealing with the FCNC problem
of ETC, namely the ideas of walking TC
and strong ETC.
Both these ideas result in the technifermion $T$ condensate
receiving a high momentum enhancement,
while the pion decay constants $F_{\pi}$ which depend on
low momentum physics (Pagels and Stokar, 1979) are almost unchanged.
This is important since the quark and lepton masses and PGB masses
depend upon the value of the condensate, while the $W,Z$ masses
depend upon $F_{\pi}$. Condensate enhancement may therefore
increase fermion masses without increasing gauge boson masses.
As we shall see,
the condensate enhancement is given by,
\beq
\psibarpsi\ \sim
 {\Lambda}_{TC}^3(M_{ETC}/{\Lambda}_{TC})^{\gamma} \label{enhancement}
\eeq
where $\Lambda_{TC}\sim 500$ GeV is the TC confinement scale,
$M_{ETC}$ is the ETC breaking scale
and $\gamma$ is the anomalous dimension of $\bar{T}T$.
The condensate is enhanced from its naive value of $\gamma=0$ to values
of up to $\gamma=1$ for walking theories or up to
$\gamma=2$ for strong ETC. It turns out that
the fermion masses $m_f$ are enhanced
by a factor of $(M_{ETC}/{\Lambda}_{TC})^{\gamma}$.
If we take $M_{ETC}\sim 500 $ TeV to avoid problems
with FCNC's then the enhancement factor is
$M_{ETC}/{\Lambda}_{TC}\sim 10^3$ and it is clear that the
enhancements in fermion masses can be large.
For example, if we take the naive value of the anomalous
dimension $\gamma=0$ then it turns out that
$m_f\sim 0.5 $ MeV. This value is fine for the
electron mass, but is clearly too small to account
for the quark masses, especially the top quark.
However if we have $\gamma=1$ then we gain a factor
of $10^3$ and $m_f\sim 500 $ MeV. This is large enough
for the strange quark mass\footnote{ Since the most
severe bound from FCNC's comes from the $K_L-K_S$
mass difference and hence $\Delta S=2$ operators,
one may argue that the relevant fermion mass
we should strive to achieve is the strange
quark mass.}.
If we wish to account for the top quark mass we need
$\gamma$ to approach its maximum value of 2,
and it turns out that this may be achieved in so-called
strong ETC theories.

As its name suggests, strong ETC theories are theories
in which the ETC gauge coupling ${\alpha}_{ETC}$
is quite strong at the scale $M_{ETC}$ at which ETC is broken.
Thus heavy ETC boson exchange can play an important role
in DEWSB at low energies.
We shall review strong ETC theories in section 3.3
where we shall see that such theories involve some degree
of fine-tuning and have light scalar bound states consisting
of tightly bound technifermion-antitechnifermion systems,
bound by short-range strongly coupled
heavy ETC boson exchange. Such theories have some
difficulties obtaining a satisfactory top-bottom mass
splitting without over-infecting the $\rho$-parameter.
One may study these theories in the four-fermion
approximation in which the exchange of ETC bosons
gives rise to effective four-technifermion operators.
\footnote{Note that it is assumed that the ETC theory
responsible gives rise to ETC bosons which couple
technifermions to technifermions. This is the case
in unitary $SU(N)_{TC}$ models for example but not
orthogonal $SO(N)_{TC}$ models.}
In this approximation the breaking of electroweak
symmetry is driven dominantly by the four-technifermion
interaction, and resembles the Nambu-Jona-Lasinio (NJL) (1961)
model.

In section 4 we discuss some of the experimental
prospects for TC models.
We begin in section 4.1
with a discussion of longitudinal gauge boson
scattering at the LHC, which is the classic signature
of the minimal TC model of section 2.1.
In section 4.2 we briefly describe some of the physics of
PGB's associated with a single family of technifermions.
This discussion is deliberately brief firstly because
it is well documented elsewhere, and secondly because
TC models with large numbers of technifermions
are somewhat under seige at the moment from
the comparison of theoretical estimates of the contribution
of radiative correction involving TC to
electroweak parameters which have been recently measured
quite accurately at LEP.
As discussed in section 4.3,
there are increasingly strong
constraints on additional fermions which couple
to the $W$ and $Z$ boson from precision electroweak measurements.
Theories which involve a complete
family of technifermions consisting of
three coloured doublets
of techniquarks $(U^{TC},D^{TC})^{r,b,g}$ plus a doublet of
technileptons $(N^{TC},E^{TC})$ are severely challenged by recent data.
There are solutions to these problems
that one can envisage so perhaps the question is not entirely
clear cut. Finally a low scale TC model
is discussed in section 4.4. In this model the TC confinement
scale $\Lambda_{TC}$ is reduced from 500 GeV to 50-100 GeV,
leading to spectacular TC signals at LEP and the Tevatron.

Another example of DEWSB
is to suppose that the top quark
is subject to some short-range strong interaction which
causes ``Cooper pairs'' of top-antitop condensates to form,
breaking the electroweak symmetry.
This is the subject of section 5.
This mechanism could also be applied to a fourth family,
providing the fourth neutrino is heavy enough so that it would
not have been produced in $Z$ decays and so would not contribute
to the accurately measured $Z$ width.
These models confront the problem of the large top quark
mass in a very direct way by doing away with TC altogether
and simply postulating some four-fermion operators
of the form
$$(\lambda / {\Lambda}^2)
(\bar{t_L}t_R\bar{t_R}t_L)$$
where $t_{L,R}$ are top quark fields, $\Lambda$ has the dimensions
of mass and is some sort of ultraviolet cut-off (which may represent
the scale at which we expect the onset of new physics such as
heavy gauge boson exchange as in the example of strong ETC above),
and $\lambda \sim 8\pi^2/3$ is a strong dimensionless coupling.
If the value of $\lambda$ is carefully adjusted the four
top quark operator becomes transmuted into a theory which
resembles the Higgs sector of the standard model
with a composite Higgs boson $H^0 \sim \bar{t}t$ coupling only
to top quarks. In section 5.3 we discuss the prospects
of a fourth family after the LEP measurement of the $Z$ width,
and in section 5.4 we shall discuss to what extent such models
predict the Higgs boson and top quark masses, due to the effects
of the so-called ``irrelevant operators'' which may
change the naive predictions based on the (quasi-)~infa-red fixed point
of these models. Finally in section 5.5 we discuss a TeV scale
model of the top quark condensate with a built-in
Glashow-Illiopoulos-Maiani
(GIM) (1970) mechanism.

Section 6 concludes the paper.

\section{Technicolour}

\subsection{The Minimal Technicolour Model}

The subject of dynamical electroweak
symmetry breaking (DEWSB) has a long and
distinguished history going back to the early work of
Nambu and Jona-Lasinio (1961).
A series of pioneering papers followed
which extended these ideas to the realm of gauge theories.
Eventually the idea of technicolour (TC) was introduced
by Susskind (1979) and Weinberg (1979)
as a mechanism for DEWSB. The early development of TC
is nicely traced in the collection or reprints by
Farhi and Jackiw (1982).

Following Susskind (1979) let us first consider a toy model
which consists of the standard model with the Higgs doublet $H$
removed, and with the fermion sector paired down to just one
family of quarks ( discarding the leptons, and
forgetting about the question of triangle
anomalies ). In this toy model the quarks $(u,d)$ remain
massless down to the QCD confinement scale $\Lambda_{QCD}\approx 250$ GeV,
whereupon the quarks and gluons become confined into hadrons
with typical masses of a GeV. What about the pions?
Susskind realised that in this toy model something very
interesting happens to the pions: they behave just like
the exact would-be Goldstone bosons of the Higgs model
and get eaten by the $W_L,Z_L$. In other words the $W,Z$
acquire mass in much the same way as in the usual Higgs mechanism.
This is such an important result which underpins much of what
follows, that it is worth understanding exactly how this
happens.

First of all let us switch off the $SU(2)_L\otimes U(1)_Y$
electroweak interactions, so that the QCD Lagrangian is invariant
under separate $SU(2)$ chiral rotations of the left-handed
quark doublet $(u,d)_L$ and the right-handed quark doublet
$(u,d)_R$, and so respects a global chiral symmetry,
\beq
SU(2)_L\otimes SU(2)_R
\eeq
together with some $U(1)$ symmetries which, although important
in some other respects, will not play a role in the following
analysis and so will be ignored. The QCD Lagrangian is
invariant under the chiral symmetry because it contains no mass
terms for the quarks. However we know from our experience
of hadron dynamics that the QCD {\em vacuum} does not respect
the full chiral symmetry which is said to be spontaneously
broken down to isospin symmetry,
\beq
SU(2)_L\otimes SU(2)_R \longrightarrow SU(2)_{L+R}
\eeq
In fact this is also an example of dynamical symmetry breaking
because the order parameter of the symmetry breaking
is the VEV of a quark-antiquark condensate,
\beq
<\bar{q}q>_{vac} \neq 0
\eeq
One can imagine a simple picture of a quark-antiquak pair
with the combined quantum numbers of the vacuum attracted by
gluon exchange, with a stronger attraction at long distances
due to the asymptotic freedom of the QCD coupling constant.
When the force of attraction becomes greater than a certain
critical value, corresponding to a certain critical
QCD coupling constant, the effective potential energy
of the ground state is lowered by producing
$q\bar{q}$ pairs. A phase transition then occurs
and it becomes energetically favourable for the vacuum
to fill up with $q\bar{q}$ pairs, and the VEV of the
bilinear operator $q\bar{q}$ becomes different from zero,
as above. It is manifestly clear that this mass-type operator
does not respect the chiral symmetry, but isospin
symmetry is preserved by the vacuum.

According to Goldstone's theorem we would expect
three massless pseudoscalars to accompany the
above chiral symmetry breaking, and phenomenologically
these are identified with the pions.
In our toy model the pions would be exactly massless
( with electroweak interactions switched off).
The following current algebra results are then exact in this limit.
The triplet of axial currents may be defined
either in terms of the quark fields $q=(u,d)$ or pion fields
$\pi^a$ as follows,
\beq
j^{\mu}_{5a}=f_{\pi}\partial^{\mu} \pi_a
= \bar{q}\gamma_{\mu}\gamma_5\frac{\tau^a}{2}q
\eeq
where $\tau^a$ ($a=1\ldots 3$) are the Pauli matrices and
$f_{\pi}$ is called the pion decay constant
because of the matrix element
\beq
<0|j^{\mu}_{5a}|\pi_b>=if_{\pi}p^{\mu}\delta_{ab}
\eeq
These results can be written equivalently as
\beqarr
j^{\mu +}_{5} & = & \sqrt{2}f_{\pi}\partial^{\mu} \pi^+
= \bar{d}\gamma_{\mu}\gamma_5u \\
j^{\mu -}_{5} & = & \sqrt{2}f_{\pi}\partial^{\mu} \pi^-
= \bar{u}\gamma_{\mu}\gamma_5d \\
j^{\mu 0}_{5} & = & f_{\pi}\partial^{\mu} \pi^0
=\frac{1}{2}( \bar{u}\gamma_{\mu}\gamma_5u - \bar{d}\gamma_{\mu}\gamma_5d)
\eeqarr
and
\beqarr
<0|j^{\mu \pm }_{5}|\pi^{\pm}> & = & \sqrt{2}if_{\pi}p^{\mu} \\
<0|j^{\mu 0 }_{5}|\pi^0> & = & if_{\pi}p^{\mu}
\eeqarr
where the charged pions are
\beq
\pi^{\pm}=\frac{\pi_1 \mp i\pi_2}{\sqrt{2}}
\eeq
and the charged currents are
\beq
j^{\mu \pm}_{5} = j^{\mu}_{51} \pm ij^{\mu}_{52}
\eeq
Experimentally $f_{\pi}\approx 93$ MeV, from the charged
pion lifetime.

The above current algebra results assume that electroweak
symmetry is switched off, so that in this limit
the pions appear in the physical spectrum as massless states.
When electroweak interactions are switched on
Goldstone's theorem ( which assumes the absence of gauge symmetries)
no longer applies. Instead we shall see that, as in the standard model,
we have a Higgs mechanism with the massless pions combining with
the massless gauge bosons to form {\em massive} gauge bosons.
Using the above results it is straightforward to show that
the kinetic terms for the quarks,
\beq
\bar{q}_LD_{\mu}\gamma^{\mu}q_L
+ \bar{u}_RD_{\mu}\gamma^{\mu}u_R
+ \bar{d}_RD_{\mu}\gamma^{\mu}d_R
\eeq
lead to the following pion-gauge boson derivative couplings,
\beq
\frac{g}{2}f_{\pi^+}W_{\mu}^+\partial^{\mu}\pi^+
+\frac{g}{2}f_{\pi^-}W_{\mu}^-\partial^{\mu}\pi^-
+\frac{g}{2}f_{\pi^0}W_{\mu}^0\partial^{\mu}\pi^0
+\frac{g'}{2}f_{\pi^0}B_{\mu}^+\partial^{\mu}\pi^0
\eeq
The derivative couplings bring down a factor of $p^{\mu}$,
the four momentum of the pion and hence that of the gauge boson
with which it mixes. In a frame in which $p^{\mu}=(E,0,0,p)$
the polarisation vectors are $\epsilon^{\mu}_T=(0,a,b,0)$
(for the transverse T vector) and $\epsilon^{\mu}_L=(c,0,0,d)$
(for the longitudinal L vector), from which it is obvious
that all the mixing takes place in the L direction only.

Using these results it is now straightforward to see
how the gauge boson masses arise.  Consider the full propagator
for the $W^{\pm}$ bosons given by summing all the diagrams
shown in figure 3. Dropping all the indices for simplicity,
the full propagator is given by the geometric series,
\beqarr
full \ \ propagator & = & \frac{1}{p^2} +
\frac{1}{p^2}(g f_{\pi^{\pm}}/2)^2 \frac{1}{p^2} + \ldots \\
& = &  \frac{1}{p^2}
\left[ 1+ \frac{(g f_{\pi^{\pm}}/2)^2}{p^2} + \ldots \right] \\
& = &  \frac{1}{p^2}
\left[ 1 - \frac{(g f_{\pi^{\pm}}/2)^2}{p^2} \right]^{-1} \\
& = & \frac{1}{p^2 - (g f_{\pi^{\pm}}/2)^2}
\eeqarr
 From above it is clear that the pole in the
gauge boson propagator is shifted from zero to $p= g f_{\pi^{\pm}}/2$
which implies that the $W^{\pm}$ has acquired a mass,
\beq
M_W= g f_{\pi^{\pm}}/2
\eeq
Note that two powers of momentum from the derivative coupling
in the numerator have cancelled with $\frac{1}{p^2}$ from
the massless pion propagator. If the pion were not exactly massless
then this cancellation would not take place, and the gauge boson
would remain massless.
A similar calculation for the $W^0$ and $B$ propagators,
including the mixing shown in figure 4, leads to the mass squared
mixing matrix for $W^0$ and $B$,
\beq
\left(
\begin{array}{ll}
M_{W_0}^2 & M_{W_0B}^2 \\
M_{W_0B}^2 & M_{BB}^2
\end{array}
\right)
= \frac{f_{{\pi}^0}^2}{4}
\left(
\begin{array}{ll}
g^2 & gg' \\
gg' & {g'}^2
\end{array}
\right)
\eeq
The eigenvalues of the matrix are
\beq
M_Z^2=(g^2+g'^2)f_{\pi^0}^2/4 , \ \ M_A^2=0
\eeq
with corresponding eigenvectors,
\beq
Z=\frac{gW_3 -g'B}{\sqrt{g^2+g'^2}}
\eeq
\beq
A=\frac{g'W_3 +gB}{\sqrt{g^2+g'^2}}
\eeq
These results are quite similar to those discussed
earlier in the standard model.
In fact the gauge boson mass ratio comes out exactly
right
\beq
M_W/M_Z=\frac{f_{\pi^{\pm}}}{f_{\pi^0}} \cos \theta_W
\eeq
since isospin symmetry guarantees that
\beq
f_{\pi^{\pm}}=f_{\pi^0}
\eeq
The only problem is that the pion decay constant $f_{\pi} \approx 93$ MeV
is too small to account for the $W,Z$ masses.
In the standard model $f_{\pi}$ is replaced by $v=246.2$ GeV.
If we could find some way of boosting $f_{\pi}$
by a factor of 2,650 then the $W,Z$ masses would have their correct
values. This is exactly what TC is: a scaled-up version
of QCD with a pion decay constant equal to $F_{\pi}=v$.

According to TC there are no elementary Higgs scalars,
instead the Higgs mechanism is implemented by the TC sector.
The TC sector consisits of a QCD-like gauge group called TC
whose coupling grows strong at around $\Lambda_{TC}\sim 500 GeV$,
together with
some technifermions which carry TC and are confined.
The TC dynamics is just a scaled up version of QCD
so that we expect chiral symmetry breaking
and light pions in TC as in QCD.
Three of the technipions will get eaten by the Higgs mechanism
and their degrees of freedom are replaced by the longitudinal (L)
components of the $W\pm,Z$\footnote{Any remaining technipions will
hopefully gain sufficient mass from explicit symmetry breaking
effects to render them too heavy to be produced.}.

We shall now discuss the simplest TC model
of Susskind (1979) and Weinberg (1979) in a little more detail.
In this minimal TC model the gauge group of the world is
\beq
SU(N)_{TC}\otimes SU(3)_C\otimes SU(2)_L\otimes U(1)_Y
\eeq
where we have added to the standard model gauge group
a new confining QCD-like gauge group $SU(N)_{TC}$ which
is asymptotically-free and confines at ${\Lambda}_{TC}\sim 500 $ GeV.
The number of technicolours $N$ is left as a free parameter
\footnote{It is conventional but not necessary to take the TC
group to be a unitary gauge group. Orthogonal TC gauge groups
are perfectly adequate alternatives.}.
The minimal set of technifermions is the anomaly-free doublet,
\beq
\begin{array}{ccl}\left( \begin{array}{c}p_L \\ m_L \end{array}
\right)^{\alpha} &\sim& (N,1,2,0)\\
{p_R}^{\alpha} &\sim& (N,1,1,1/2)\\
{m_R}^{\alpha} &\sim& (N,1,1,-1/2)
\end{array}
\eeq
where ${\alpha}=1\ldots N$ labels TC.
Each techniquark carries TC but not ordinary colour, and the left-handed
techniquarks form an $SU(2)_L$ doublet just like ordinary quarks.
The electric charge generator is given by $Q=T_{L3} + Y$ so that
the plus ($p$) and minus ($m$) techniquarks have charges given by
$Q=\pm 1/2$,respectively.
The electric charges sum to zero, a requirement of
anomaly freedom.
Needless to say the quarks and leptons defined earlier
do not carry technicolour.

There is an obvious parallel between the technidoublet above
and the doublet of lightest quarks in QCD, as shown in Table 1.

\begin{table}
\caption{
Comparison of minimal TC and QCD with one quark doublet.}
\begin{tabular}{|c|c|} \hline\hline
{\em QCD} & {\em TC} \\ \hline
$SU(3)_C$            &       $SU(N)_{TC}$ \\
Quarks               &         Techniquarks\\
$\left( \begin{array}{c}u \\ d \end{array} \right)$
&
$\left( \begin{array}{c}p \\ m \end{array} \right)$ \\
$<\bar{u}u+\bar{d}d>\sim {\Lambda}_{QCD}^3$
&
$<\bar{p}p+\bar{m}m>\sim {\Lambda}_{TC}^3$  \\ \hline
\multicolumn{2}{|c|}{$SU(2)_L\otimes SU(2)_R\rightarrow SU(2)_{L+R}$}
\\
\multicolumn{2}{|c|}{Chiral Symmetry $\rightarrow$
(techni-)isospin} \\  \hline
3 QCD pions              &         3 Technipions     \\
$f_{\pi}^{\pm}=f_{\pi}^{0}=93 MeV$    &
$F_{\pi}^{\pm}=F_{\pi}^{0}=246 GeV$    \\
${\Lambda}_{QCD}\approx 200 MeV$   &
${\Lambda}_{TC}\approx 500 GeV$   \\ \hline \hline
\end{tabular}
\end{table}

In the idealised world of table 1 consisting of a doublet
of massless quarks and a doublet of massless techniquarks,
the 3 pions and 3 technipions are massless up to electroweak
effects. When $SU(2)_L\otimes U(1)_Y$
forces are switched on, 3 of the pions will be eaten by the Higgs
mechanism and 3 pions will remain in the spectrum as physical states.
The 3 eaten pions are mainly technipions, and the 3 physical pions
are mainly QCD pions,
\beqarr
| eaten \; pion> & = &
 \frac{F_{\pi}| technipion>+f_{\pi}| QCD \; pion>}
{\sqrt{F_{\pi}^2 + f_{\pi}^2}}  \\ \nonumber
| physical \;  pion> & = &
 \frac{F_{\pi}|QCD \; pion>-f_{\pi}| technipion>}
{\sqrt{F_{\pi}^2 + f_{\pi}^2}}.
\eeqarr
In the approximation $F_{\pi} \gg f_{\pi}$, the gauge boson
masses are given by,
\beq
M_{W^{\pm}}^2=\frac{g^2F_{{\pi}^{\pm}}^2}{4}
\eeq
\beq
\left(
\begin{array}{ll}
M_{W_0}^2 & M_{W_0B}^2 \\
M_{W_0B}^2 & M_{BB}^2
\end{array}
\right)
= \frac{F_{{\pi}^0}^2}{4}
\left(
\begin{array}{ll}
g^2 & gg' \\
gg' & {g'}^2
\end{array}
\right)
\eeq
In the limit of exact techni-isospin $SU(2)_{L+R}$ we have $F_{{\pi}^+}=
F_{{\pi}^0}$ and these results are identical to those of the standard
model if we identify $F_{\pi}=v$ where $v$ is the vacuum expectation
value (vev) of the Higgs doublet, and $g$ and $g'$ are the $SU(2)_L$
and $U(1)_Y$ gauge couplings.
When the mass-squared matrix is diagonalised we obtain a massive $Z^0$
gauge boson and a massless photon.
The gauge boson masses are thus given by
\beq  M_{W^{\pm}}=\frac{gF_{{\pi}^{\pm}}}{2}       \eeq
\beq  M_{Z^{0}}=\frac{gF_{{\pi}^{0}}}{2 \cos {\theta}_W}.  \eeq
We arrange $F_{\pi}=246$ GeV in order to give masses of the correct
magnitude. This is the reason why we want a large confinement
scale $\Lambda_{TC}\sim  500 $ GeV.

\subsection{Extended Technicolour}

The minimal TC model just described provides a neat way of
breaking $SU(2)_L\otimes U(1)_Y$ without using the Higgs doublet
$H$. In providing masses for the
$W$ and $Z$ TC is unsurpassed in its elegance.
The successful $W/Z$ mass ratio is
a consequence of the global $SU(2)_{L+R}$
techni-isospin symmetry (the analogue of ordinary isospin
symmetry in QCD).
However $H$ is not only is responsible for gauge boson masses,
but also provides fermion masses via the Yukawa couplings which
break the chiral symmetries of the fermion sector.
So far we have not mentioned how quark and lepton masses
may be achieved in TC theories. Now we shall discuss this question.

In the standard model
the Higgs doublet has Yukawa couplings to fermions $f$
of the form $\bar{f}_{iR}H^{\dagger}f_{jL}$.
In minimal TC the Higgs doublet $H$ is replaced by
the technifermion doublet $T=(p,m)$ which have no couplings
to quarks and leptons, except through chirality conserving
gauge interactions. Therefore quark and lepton masses are
not generated in the minimal TC model.
It is at this point that many people choose to
give up on TC theories
since they do not possess the economy of the single Higgs
doublet\footnote{Higgs kills two birds with one stone by providing
both gauge boson masses and fermion masses.}.
However TC enthusiasts claim that the economy of the Higgs doublet
is illusory, and in reality the arbitrary Yukawa couplings
of the standard model just emphasise the fact
that we have absolutely
no understanding of the origin of fermion masses.
The fact that the fermion masses come out to be zero in minimal
TC could even be construed as a good thing, since this simple
fact forces one to introduce theories of fermion masses
from the outset.
The fact that it will turn out to be exceedingly difficult
to correctly account for the fermion mass spectrum, without
running into phenomenological difficulties of one sort or another,
can again be interpreted either positively or negatively.
TC sceptics regard these difficulties as hints that Nature
does not like TC theories. TC believers regard these difficulties
as challenges which must and can be overcome.
There are so many approaches to the fermion mass problem
in TC theories that this subject has become model-dependent in the extreme.
In this sub-section
we shall concentrate on the general strategy known as
extended technicolour, bearing in mind that there are other approaches
to the problem. Later on we shall consider some of the other approaches
(e.g. the top quark condensate approach in section 5).

One solution to the problem of fermion masses in TC is to
extend the TC sector in such a way as to allow
the technifermons to talk to the fermions (quarks and leptons)
and so allow the dynamically generated technifermion mass
to be fed down to the fermions by some sort of radiative correction.
This strategy is called extended technicolour (ETC)
and was first discussed by Dimopoulos and Susskind (1979)
and Eichten and Lane (1980).
The general strategy of ETC theories is clear: one embeds the TC
gauge group $G_{TC}$ into a larger ETC gauge group $G_{ETC}$
which is broken
{\em somehow} at a scale $M_{ETC}$ down to $G_{TC}$,
\beq
G_{ETC}
 \begin{array}{c} M_{ETC} \\ \longrightarrow \end{array}
G_{TC}\otimes \cdots
\eeq
where $M_{ETC}>{\Lambda}_{TC}\sim 500 GeV$.
The heavy ETC gauge bosons
of mass $M_{ETC}\sim 1-1000$ TeV, corresponding to the broken
ETC generators, can in general can couple
fermions (f) to technifermions (T) , fermions to fermions, and
technifermions to technifermions,
as shown in figure 5. The ETC
coupling of
technifermions
to fermions, allows the radiative diagram in figure 6.
This diagram allows the feed-down of the
dynamically generated technifermion mass to the quarks and leptons,
and leads to a fermion mass $m_f$ crudely given by,
\beq
m_f \sim \frac{<\bar{T}T>_{M_{ETC}}}{M_{ETC}^2} \label{mf}
\eeq
where a naive estimate of the technifermion condensate,
based on simple dimensional analysis is,
\beq
<\bar{T}T>_{M_{ETC}} \sim {{\Lambda}_{TC}}^3. \label{condensate1}
\eeq
We can improve on the naive estimates above of the fermion mass in ETC,
and in section 3 we shall do exactly that. But for now we shall be content
with the simple estimates above.

However according to figure 5
the ETC bosons also couple fermions to
fermions which leads to flavour changing neutral currents (FCNC).
The most severe constraint (Eichten and Lane, 1980) comes
from $\Delta S=2$ operators $\sim (1/{M_{ETC}}^2)\bar{s}d \bar{s}d$
which mediate $\bar{s}d \leftrightarrow \bar{d}s$ mixing
due to the diagram in figure 7.
The $K_L-K_S$ mass difference leads to the constraint
$M_{ETC}>500 TeV$. This in turn leads to a bound on the fermion
mass $m_f<0.5 MeV$ from equations \ref{mf}, \ref{condensate1}.
The FCNC problem is simply
that such fermion masses are unrealistically small.
Such problems arise in part because ETC theories have
set themselves the ambitious task of explaining all the quark
and lepton masses and quark mixing angles without the aid
of elementary scalar fields.

Apart from the FCNC problem, ETC theories face the problem
of ensuring that there are no massless
Goldstone bosons (GB's) in the spectrum.
In the minimal TC model, in the limit that electroweak interactions
were switched off, there are exactly three massless GB's.
When electroweak interactions are switched on these get eaten by
the $W,Z$, according to the Higgs mechanism, leaving no GB's
in the physical spectrum. However many ETC models give rise to a
low-energy effective TC theory which involves more than one
doublet of technifermions, and hence has a larger chiral symmetry
than in the minimal model. When this larger chiral symmetry is broken,
there will be more than three broken generators leading to more
than three GB's, according to Goldstone's theorem.
The solution to this potential disaster is to ensure that the
original chiral symmetry of the TC theory is not exact, but is
in fact explicitly broken in some way, so that we do not end up
with exactly massless GB's but instead rather light pseudo-Goldstone bosons
(PGB's). The PGB's of course may be problematic unless they are
sufficiently heavy, but at least we have a strategy for solving
this difficulty.

The simplest example of PGB's
are the ordinary pions of QCD.
The pions are not exactly massless since the original
chiral symmetry is explicitly broken by two effects.
Firstly the quarks $u,d,\ldots$ have explicit masses
(sometimes called current masses or intrinsic masses)
which clearly do not respect the chiral symmetry.
Secondly the quarks carry electric charges which violate
the isospin symmetry. The mass difference between the charged
and neutral pion is entirely due to electromagnetic effects.
The mass of the neutral $\pi^0$ on the other
hand is totally due to the effect of the explicit
quark masses which break the chiral symmetry explicitly.
The value of the pion masses may be estimated
using either current algebra techniques or
chiral lagrangian techniques which are equivalent to the
current algebra techniques but much nicer to work with.
Chiral lagrangians are discussed by Georgi (1984).
The well known result is that the neutral pion mass squared
varies as,
\beq
m_{\pi^0}^2 \sim (m_u + m_d)f_{\pi}
\eeq
where $m_u,\ m_d$ are the explicit quark masses,
and as before $f_{\pi}=93$ MeV is the pion decay constant.
The difference between the squares of the masses of the
charged and neutral pions varies as,
\beq
m_{\pi^{\pm}}^2-m_{\pi^0}^2 \sim \alpha f_{\pi}^2
\label{piondiff}
\eeq
where $\alpha=1/137$ is the fine structure constant.

In TC theories with more than one doublet of technifermions,
the PGB's which carry colour or electroweak quantum numbers
will receive masses which may be estimated by scaling
up the charged-neutral pion mass difference in equation \ref{piondiff}
by using $F_{\pi}=246$ GeV and replacing $\alpha$ by the
appropriate coupling factor. The electrically neutral PGB's may
or may not receive mass from electroweak effects.
However in ETC theories there are also ETC gauge boson effects
to take into account, as these may also explicitly violate
the chiral symmetry of the low-energy TC theory. The precise
details will depend on the particular ETC model under consideration,
and in fact the requirement of no extra GB's will put powerful
constraints on the ETC theory as emphasised by Eichten and Lane (1980).
Assuming the ETC theory has the correct interactions to break
the chiral symmetry of the TC theory, the resulting masses
from ETC effects are estimated to be
\beq
m_{P}^2 \approx \frac{<\bar{T}T>}{F_{\pi}^2M_{ETC}} \label{ETCpi}
\eeq
where $P$ represents a PGB of TC. Such mass contributions
may be the only source of mass for the neutral PGB's,
and may be an important source of mass for the electrically charged
but colour singlet PGB's, which must be sufficiently
heavy to not have been produced at the LEP $e^+e^-$ collider.

\subsection{A simple ETC model}

So far we have been deliberately vague about the details of the ETC theory.
This is because, as mentioned, ETC theories are highly model dependent.
There are about as many different ETC models as there are people
who have ever worked on ETC. There is certainly no leading candidate
ETC model, and all ETC models that have ever been proposed has
difficulties of one sort or another. Nevertheless it is important
to discuss explicit detailed models, since only then can one gain
an appreciation of what the issues are and what the kinds of problems
are that one may expect to face. With this
in mind I shall now discuss a very basic prototype ETC model
first introduced by the present author in 1989.
The ETC model involves a complete family of technifermions,
i.e. a set of technifermions with the quantum numbers of a complete
quark and lepton family, but which carry an additional TC quantum number.
It is a common feature of ETC models that they yield a
single complete techni-family in the low energy TC theory.
However this is by no means a necessary consequence of ETC,
and there are many other ETC models which do not yield a techni-family.
Another feature of the model we are about to descibe is that it
is a multi-scale ETC model. In other words, the ETC gauge group
does not break down to TC at a single energy scale, but
rather it breaks sequentially over a hierarchy of scales.
The hierarchy of ETC scales is a simple reflection of the
hierarchy of fermion masses between families.

Consider the following ETC theory (King 1989a),
\beq
SO(10)_{ETC}\otimes SU(3)_C \otimes SU(2)_L \otimes U(1)_Y
\eeq
\beq
\begin{array}{ccl}
{Q_L^{ETC}}^{\alpha} =
\left( \begin{array}{c}U_L^{ETC} \\ D_L^{ETC} \end{array}
\right)^{\alpha} &\sim& (10,3,2,1/6)\\
{U_R^{ETC}}^{\alpha} &\sim& (10,3,1,2/3)\\
{D_R^{ETC}}^{\alpha} &\sim& (10,3,1,-1/3)\\
{L_L^{ETC}}^{\alpha} =
\left( \begin{array}{c} N_{L}^{ETC} \\ E_L^{ETC} \end{array}
\right)^{\alpha} &\sim& (10,1,2,-1/2)\\
{E_R^{ETC}}^{\alpha} &\sim& (10,1,1,1)\\
{N_R^{ETC}}^{\alpha} &\sim& (10,1,1,0)
\end{array}
\eeq
where $\alpha=1 \ldots 10$ is an $SO(10)_{ETC}$ index.
Apart from the index $\alpha$ these fermions have the
quantum numbers of a family of quarks and leptons.
Let us assume that the ETC symmetry is broken sequentially
at three ETC scales as shown below,
\beq
SO(10)_{ETC}\stackrel{M_1^{ETC}}{\rightarrow}
SO(9)_{ETC}\stackrel{M_2^{ETC}}{\rightarrow}
SO(8)_{ETC}\stackrel{M_3^{ETC}}{\rightarrow}
SO(7)_{TC}
\eeq
The final symmetry group $SO(7)_{TC}$ is assumed to be unbroken,
and is identified as a TC gauge group which confines at the
TC scale $\Lambda_{TC}$. At each stage of symmetry breaking,
the ETC representation decomposes into a smaller ETC
representation plus a TC singlet, according to,
\beq
10 \stackrel{M_1^{ETC}}{\rightarrow} 9 \oplus 1
\eeq
\beq
9 \stackrel{M_2^{ETC}}{\rightarrow} 8 \oplus 1
\eeq
\beq
8 \stackrel{M_3^{ETC}}{\rightarrow} 7 \oplus 1
\eeq
Each of the TC singlets so produced is identified with a
family of quarks and leptons. Thus the low-energy TC theory
consists of three families of
quarks and leptons plus one techni-family,
\beq
SO(7)_{TC}\otimes SU(3)_C \otimes SU(2)_L \otimes U(1)_Y
\eeq
\beq
\begin{array}{ccl}
{Q_L^{TC}}^{\alpha} =
\left( \begin{array}{c}U_L^{TC} \\ D_L^{TC} \end{array}
\right)^{\alpha} &\sim& (7,3,2,1/6)\\
{U_R^{TC}}^{\alpha} &\sim& (7,3,1,2/3)\\
{D_R^{TC}}^{\alpha} &\sim& (7,3,1,-1/3)\\
{L_L^{TC}}^{\alpha} =
\left( \begin{array}{c}N_{L}^{TC} \\ E_L^{TC} \end{array}
\right)^{\alpha} &\sim& (7,1,2,-1/2)\\
{E_R^{TC}}^{\alpha} &\sim& (7,1,1,1)\\
{N_R^{TC}}^{\alpha} &\sim& (7,1,1,0)
\end{array}
\eeq
\beq
\begin{array}{ccl}
{Q_L}^{i} = \left( \begin{array}{c}U_L \\ D_L \end{array}
\right)^{i} &\sim& (1,3,2,1/6)\\
{U_R}^{i} &\sim& (1,3,1,2/3)\\
{D_R}^{i} &\sim& (1,3,1,-1/3)\\
{L_L}^{i} = \left( \begin{array}{c}\nu_{L} \\ E_L \end{array}
\right)^{i} &\sim& (1,1,2,-1/2)\\
{E_R}^{i} &\sim& (1,1,1,1)\\
{\nu_R}^{i} &\sim& (1,1,1,0)
\end{array}
\eeq
where $\alpha=1 \ldots 7$ is an $SO(7)_{TC}$ index,
and $i=1 \ldots 3$ labels the three families of quarks and leptons.

There is no Higgs doublet in this model, and electroweak
symmetry is broken by the TC condensates,
\beqarr
<\bar{U}^{TC}_LU^{TC}_R> & \ne & 0 \\
<\bar{D}^{TC}_LD^{TC}_R> & \ne & 0 \\
<\bar{E}^{TC}_LE^{TC}_R> & \ne & 0 \\
<\bar{N}^{TC}_LN^{TC}_R> & \ne & 0
\eeqarr
Quark and lepton masses arise from the graphs in figure 8.
The resulting fermion masses according
equation \ref{mf} are thus,
dropping the ETC and TC superscripts,
and the L and R subscripts, for brevity,
\beqarr
m_t & \sim & \frac{<\bar{U}U>}{M_{3}^2}, \
m_c \sim \frac{<\bar{U}U>}{M_{2}^2}, \
m_u \sim \frac{<\bar{U}U>}{M_{1}^2} \\
m_b & \sim & \frac{<\bar{D}D>}{M_{3}^2}, \
m_s  \sim  \frac{<\bar{D}D>}{M_{2}^2}, \
m_d  \sim  \frac{<\bar{D}D>}{M_{1}^2} \\
m_{\tau} & \sim & \frac{<\bar{E}E>}{M_{3}^2}, \
m_{\mu} \sim  \frac{<\bar{E}E>}{M_{2}^2}, \
m_e  \sim  \frac{<\bar{E}E>}{M_{1}^2} \\
m_{\nu_{\tau}} & \sim & \frac{<\bar{N}N>}{M_{3}^2}, \
m_{\nu_{\mu}}  \sim \frac{<\bar{N}N>}{M_{2}^2}, \
m_{\nu_e}  \sim  \frac{<\bar{N}N>}{M_{1}^2}
\eeqarr
The ETC bosons all commute with the standard model gauge group
and so are electrically and colour neutral.
The mass scale of each quark and lepton family
is set by the ETC scale at which that particular family
was born as a TC singlet. Thus the family mass hierarchy
may be associated with the three ETC scales $M_i$.
Within a particular family, the large mass splittings
which may occur are much harder to account for.
For instance the top-bottom mass splitting
is at odds with our natural expectation that
\beq
<\bar{U}U>=<\bar{D}D>=<\bar{E}E>=<\bar{N}N>\neq 0
\eeq
Indeed the first and last equalities
are necessary in order to preserve techni-isospin
symmetry which is vital for the correct $W/Z$ mass
relations.
The techniquark condensates $<\bar{U}U>=<\bar{D}D>$
may be allowed to differ from the technilepton condensates
$<\bar{E}E>=<\bar{N}N>$, for example on the grounds
that techniquarks carry colour as well as technicolour,
whereas technileptons only carry technicolour,
so the bottom-tau splitting may in principle be explainable,
as we mention in section 3.2.
Also the smallness of the neutrino masses may be accounted for
by assuming that the original right-handed ETC-neutrino fields
$N^{ETC}_R$ are given a Majorana mass.
This will then result in both the right-handed technineutrinos
$N^{TC}_R$ and ordinary right-handed neutrinos
$\nu_{iR}$ having the same Majorana mass.
After the technineutrino condensates form
small physical neutrino masses may then result as a consequence of
a see-saw mechanism. Majorana technineutrino masses also lead to negative
contributions to the S and T parameters,
as we discuss in section 4.3.
Note that Majorana masses are only allowed in the context of
orthogonal gauge groups, e.g. if we had chosen
the ETC gauge group to be $SU(10)_{ETC}$ rather than
$SO(10)_{ETC}$ then we would not have been allowed
to assume any Majorana masses \footnote{This is because
for $SU(N)$ groups the product $N\otimes N$ does
not contain a singlet whereas for $SO(N)$ groups
it does. This in turn is because for $SU(N)$
$N$ is a complex representation,
while for $SO(N)$, $N$ is real.}.
Finally quark mixing angles are all zero in this model.

Clearly the model has some problems, but is useful
as a simple example of an ETC model.
As shown by King (1989b) and King and Mannan (1992)
it is possible to
fix-up the model by introducing two ETC gauge groups
\beq
SO(10)_{ETC1}\otimes SO(10)_{ETC2}
\eeq
The low-energy TC theory is the same, but the undesirable
fermion mass relations may all be cured. The PGB spectrum,
and the question of FCNC's in this model are also
discussed in the above paper. The analysis of King and Mannan
(1992) involves a numerical solution to the full set
of coupled gap equations and involves the techniques to be
discussed in section 3.

We should emphasise that there
are many other ETC models in the literature
which use the condensate enhancement techniques
about to be discussed. For example
Guidice and Raby (1992) have attempted to
construct a grand unified ETC model.
By contrast Sundrum (1993) has constructed an ETC model
valid up to 150 TeV in which electroweak symmetry
is broken by a single technidoublet.
Recently Appelquist and Terning (1994) have attempted
to construct an ETC model in which the ETC gauge group
is broken by a tumbling mechanism.
Tumbling is essentially the statement that chiral gauge
groups can become strong and form condensates in the most attractive
channel (MAC) which lead to the gauge group being broken
to a smaller gauge group (Dimopoulos {\it et al} 1980).
This is a nice idea because it means that one does not
have to re-introduce scalars in order to break the ETC gauge group.

\section{Condensate Enhancement}

\subsection{Schwinger-Dyson Gap Equation}

We have seen in equation \ref{mf} that the fermion mass is proportional
to the technifermion condensate
\psibarpsi\ and that consequently if one wishes to enhance
fermion masses one must enhance the condensate
as in equation \ref{enhancement}.
Similarly, the ETC contribution to PGB masses in equation \ref{ETCpi}
is also proportional to the condensate, and may be similarly enhanced.

Fermion masses arise from the feed-down of the dynamically
generated technifermion mass as shown in figure 6.
One may approximate the loop calculation in figure 6 by writing the
ETC gauge boson propagator as $g^{\mu \nu}/(-M_{ETC}^2)$
(i.e. neglect the momentum $p$ dependence compared to the
heavy boson mass $M_{ETC}$) and then introduce an ultraviolet
cut-off $M_{ETC}$ into the Euclidean space loop integral.
The result is, after performing the angular integrations,
\beq
m_f\approx \frac{g_{ETC}^2}{M_{ETC}^2}
\frac{N}{4{\pi}^2}
\int_{0}^{M_{ETC}^2}
dp^2p^2 \frac{\Sigma (p^2)}{p^2 + {\Sigma}^2(p^2)}
\eeq
where $g_{ETC}$ is the ETC gauge coupling constant evaluated at the
scale $M_{ETC}$, $N$ is the number of technicolours carried by the
technifermions, and $\Sigma(p^2)$ is the running technifermion mass
in Euclidean space. $\Sigma(p^2)$ is the self-generated mass of the
technifermions and is the direct analogue of the dynamically
generated quark mass in QCD. In QCD we have good reasons for believing
that $\Sigma(0) \approx 300 MeV$ ( a third of the proton mass)
and that $\Sigma(p^2)$ decreases for large $p^2$ like
$1/p^2$ (times a log) (Lane 1974, Politzer 1976).
It is conventional to define the condensate as,
\beq
\psibarpsi\ =
\frac{N}{4{\pi}^2}
\int_{0}^{M_{ETC}^2}
dp^2p^2 \frac{\Sigma (p^2)}{p^2 + {\Sigma}^2(p^2)} \label{condensate2}
\eeq
so that the fermion mass is given by
\beq
m_f\approx \frac{g_{ETC}^2}{M_{ETC}^2}
\psibarpsi\
\eeq
Since gauge couplings are typically of order unity this result
agrees with equation \ref{mf}.
Futhermore if we assume that $\Sigma(0) \approx {\Lambda}_{TC}$
and falls like $1/p^2$ then up to factors and a log
we reproduce the dimensional analysis
estimate of the condensate in equation \ref{condensate1}.
It is important to know how the technifermion dynamical
mass function $\Sigma(p^2)$ falls at large momentum
in order to estimate the condensate reliably.
For example if $\Sigma(p^2)\sim {\Lambda}_{TC}^2/p$
then the condensate goes as,
$\psibarpsi\ \sim {\Lambda}_{TC}^2{M_{ETC}}$ and is enhanced
by a factor of $(M_{ETC}/ {\Lambda}_{TC})$, corresponding
to the enhancement mentioned
in equation \ref{enhancement} with $\gamma=1$.
This is precisely what happens in walking TC theories.
In order to discuss this effect quantitatively we need to
understand how to estimate the behaviour of $\Sigma(p^2)$.
We shall shortly introduce a gap equation which will
enable $\Sigma(p^2)$ to be determined. But before doing so,
we shall write down the Pagels-Stokar (1979) formula
for the pion decay constant $F_{\pi}$,
\beq
F_{\pi}^2=\frac{N}{(2\pi^2)}
\int_{0}^{M_{ETC}^2}dp^2p^2 \Sigma(p^2)
\frac{\Sigma(p^2)-\frac{1}{2} p^2 \frac{d\Sigma(p^2)}{dp^2} }
{(p^2+\Sigma(p^2)^2)^2} \label{fpi}
\eeq
It is clear that while the condensate in equation \ref{condensate2}
is sensitive
to the high $p$ region of the integration, the pion decay constant
in equation \ref{fpi} is relatively insensitive to this region.
Thus quark and lepton masses may be enhanced without changing the
$W$ and $Z$ masses very much.

The technifermion mass function $\Sigma(p^2)$ may be
estimated from the Schwinger-Dyson gap equation in
ladder approximation. We shall follow
the approach of Appelquist {\it et al} (1986, 1987, 1988a)
where the gap equation is given by,
\beq
\Sigma(p^2) = \frac{3C_2(R)}{4\pi}
\int_{0}^{M_{ETC}^2}
dk^2k^2 \frac{{\alpha}_{TC}(max(k^2,p^2))}{max(k^2,p^2)}
\frac{\Sigma (k^2)}{k^2 + {\Sigma}^2(k^2)} \label{gap}
\eeq
where $C_2(R)$ is the quadratic Casimir of the technifermions
in the TC representation\footnote{If the technifermions are
in the fundamental representation of $SU(N)$ then $R=N$.} $R$,
${\alpha}_{TC}(max(k^2,p^2)$ is the running TC coupling
evaluated at the maximium of $k^2$ and $p^2$.
The gap equation above is valid in the Landau gauge where the
fermion wavefuncion renormalistion $Z(p^2)=0$.
The angular integrations have been performed in Euclidean space,
and the ETC scale has been inserted as an ultraviolet cut-off (physically
it corresponds to the onset of new physics beyond TC, i.e. ETC).
The above gap equation is represented diagrammatically
in figure 9.

The justification for the use of the gap equation is discussed by
Appelquist and Wijewardhana (1987).
One should be aware of the well known criticisms of the
use of this equation to estimate such quantities as the condensate.
Essentially the gap equation is an uncontrolled truncation
of the full Schwinger-Dyson gap equations.
It is valid only in one gauge and so is not
manifestly gauge invariant. Finally it involves a running coupling
constant inside the integrand, which purists may object to.
Nevertheless we shall use the gap equation to give us insights
into the dynamics which are not possible to achieve without it.
Despite all its failings, it is undoubtedly
a really useful equation.

In order to solve the gap equation we need some ansatz for the
running TC coupling constant ${\alpha}_{TC}(p^2)$.
Since the lower limit of integration extends into the non-perturbative
region, we need to model the coupling in this region somehow.
The ansatz used is the following.
The technifermions are assumed to condense at a scale
$\mu=2\Sigma(0)$ so that, for $p>\mu$, ${\alpha}_{TC}$ runs with
all the fermions present, while for ${\Lambda}_{TC}<p<\mu$
it runs with no fermions present. The scale ${\Lambda}_{TC}$
represents the TC confinement scale, below which the coupling
is taken to be some constant value ${\alpha}_{TC}^0$.
By studying the effective potential (Peskin 1982)
one may estimate that the critical coupling constant
at which the technifermions condense is ${\alpha}_{TC}^c
=\pi/3C_2(R)$. If one takes this estimate seriously then one
may arrange that in the above ansatz
${\alpha}_{TC}(\mu)={\alpha}_{TC}^c$.
With the ansatz for ${\alpha}_{TC}$ specified then one may
proceed to solve the gap equation by recursive numerical techniques.
The results turn out to be remarkably sensitive
to the behaviour of the coupling constant in the high momentum
region. Above the chiral symmetry breaking scale $\mu=2\Sigma(0)$
the TC coupling evolves according to the one loop result,
\beq
{\alpha}_{TC}(p^2>\mu^2)=\frac{{\alpha}_{TC}(\mu^2)}
{1 + b{\alpha}_{TC}(\mu^2)\log(p^2/\mu^2)}
\eeq
where
\beq
b=(\frac{11}{3} C_2(G) - \frac{4}{3} T(R)n_f)
\eeq
and $C_2(G)$ is the quadratic Casimir of the adjoint representation
of the technigluons $G$, $T(R)$ is the index of the technifermion
representation $R$
and $n_f$ is the number of Dirac technifermons.
More generally the beta function is given by,
\beq
{\beta}_{TC}=p\frac{d {\alpha}_{TC}}{dp}
= -b{\alpha}_{TC}^2 - c{\alpha}_{TC}^3 \ldots \label{beta}
\eeq

\subsection{Walking Technicolour}

Holdom's original observation back in 1981 was that
if the TC theory was born at a fixed point ${\alpha}_{TC}\neq 0$,
${\beta}_{TC}=0$ then the condensate could be enhanced
as in equation \ref{enhancement} with $\gamma \sim 1$.
This assumption of
a fixed point TC theory was later considered by
Yamawaki {\it et al} (1986a, 1986b).
Condensate enhancement (without $F_{\pi}$ being changed)
is also possible for asymptotically free (ASF) theories
in which $\beta_{TC}\approx 0$ and the coupling is slowly
running (walking) (Holdom 1985, Appelquist {\it et al} 1986, 1987).
Walking TC theories rely on either
a large number $n_f\gg 2$ of technifermions in the
fundamental representation $R=N$ (type C theories)
or $n_f=2$ technifermions in a very large representation
$R\gg N$ (type A theories) , or some other combination
(type B theories), in order to arrange that $b\approx 0$.
Care must be taken to ensure that $b>0$ (to maintain ASF),
and $c{{\alpha}_{TC}^c}^3< b{{\alpha}_{TC}^c}^2$
(for good convergence of the series in equation \ref{beta}).
Numerical solutions of equation \ref{gap}
are depicted in figure 10.
These numerical solutions for $\Sigma(p^2)$ are inserted
into equation \ref{condensate2} in order to estimate the condensate.
Detailed numerical studies (Appelquist {\it et al} 1988a,
King and Ross 1989)
confirm the simple
depictions in figure 10, and demonstrate that the values of $F_{\pi}$
obtained from the Pagels-Stokar result are insensitive
to high-p physics.

The gap equation may also receive contributions
from other sources as shown in figure 11.
Each of these diagrams in principle must
be added to the right-hand side of figure 9.
The next order technigluon corrections in figure 11a
beyond the ladder approximation have been studied
by Appelquist {\it et al} (1988b) and Holdom (1988a)
and tend to decrease the critical coupling $\alpha_{TC}^c$
by a few per cent for type A walking theories, and by
about $20\%$ for type C walking theories.
The question of gauge dependence and higher order corrections
has also been studied by Kamli and Ross (1992), who conclude that
there are quite large uncertainties in the estimation of the
condensate.
Another important correction for techniquarks $Q$ (absent for
technileptons $L$) is QCD gluon exchange (Holdom, 1988b)
in figure 11b.
This can result in condensate splitting
$<\bar{Q}Q>/<\bar{L}L>\sim 10$, which may split the quark and
lepton masses by a similar factor if one assumes that quarks receive
mass from techniquarks, and leptons from technileptons
as in the simple model of section 2.3 for example.
ETC boson exchange corrections in figure 11c
have been studied
in the four-fermion approximation by Appelquist {\it et al} (1989)
and Holdom (1989), and in the full theory by King and Ross (1990)
and give important condensate enhancement if the ETC
coupling constant $\alpha_{ETC}$ is sufficiently strong.
This effect (called strong ETC) will be discussed in section 3.3.
Finally another important contribution to the technifermion condensate
is top quark exchange, mediated by ETC bosons as shown in figure 11d
and discussed by Appelquist and Shapira (1990)
and King and Mannan (1992).
This effect is interesting because the top quark is treated
as a dynamical fermion on the same footing as the techniquarks.
For example, in the orthogonal ETC model of King and Mannan (1992)
there are no
ETC bosons which connect techniquarks to techniquarks,
but there are ETC bosons which connect top quarks to techniquarks.
These ETC bosons allow the techniquark mass to be fed-down
to the top quark mass, but because the top quark
is so heavy the top quark mass also contributes to the
techniquark mass by a similar diagram.
Thus there is a sort of feed-back effect which can enhance
both the techniquark and the top quark masses.
This is in fact a simple example of a top quark condensate,
about which we shall say more in section 5.

\subsection{Strong Extended Technicolour}

Walking TC can at best lead to $\Sigma(p^2)\sim 1/p$ and hence
$\gamma \approx 1$. As discussed in the introduction this amount
of condensate enhancement is welcome but insufficient to account
for heavy quark masses, especially the top quark mass.
The ultimate condensate enhancement arises from $\Sigma$ which is
flat out to $M_{ETC}$, corresponding to $\gamma \sim 2$
in equation \ref{enhancement}.
Such condensate enhancements would be certainly sufficient to account
for the top quark mass, if they can be achieved
without ill effects.

As mentioned, condensate enhancements up to $\gamma =2$
can result from the idea of strong ETC
(Appelquist {\it et al} 1989,
Holdom 1989, King and Ross 1990).
What is strong ETC? It is a correction to the gap equation
coming from the diagram in figure 11c. Thus there is a second
term on the rhs of equation \ref{gap} which is similar to the original term
but with $\alpha_{TC}$ replaced by $\alpha_{ETC}$ , the Casimir
replaced by a different Casimir associated with the embedding
of $G_{TC}$ into $G_{ETC}$  and
with $max(k^2,p^2)$ replaced by $max(k^2,p^2,M_{ETC}^2)$
which approximates the exchange of the heavy ETC boson as a step
function, where the upper limit of integration now extends
to infinity.
Strongly coupled heavy ETC gauge boson exchange was first studied
by King and Ross (1990).
The effect of heavy ETC boson exchange may also be approximated
by making a four fermion approximation and cutting off the upper limits
of integration at $M_{ETC}$,
and this approximation was originally made
by Appelquist {\it et al} (1989) and Holdom (1989).
In all these approaches the physical effect is the same:
for sufficiently strong ETC couplings $\alpha_{ETC}$
the exchange of the heavy boson can dominate the physics of
electroweak symmetry breaking, with technigluon exchange
being of secondary importance. The result of this is that
electroweak symmetry is broken dominantly by physics associated
with a high energy scale $M_{ETC}$. For sufficiently large
$\alpha_{ETC}$ the model will resemble the NJL model
and all technifermion masses $\Sigma$ will be of order $M_{ETC}$.
This is undesirable since it implies that the $W,Z$ masses
will similarly be $0(M_{ETC})$.

We are interested in values of $\alpha_{ETC}$ which are
strong but not too strong. What happens to $\Sigma$ as we
increase the value of $\alpha_{ETC}$ from zero up to
some strong value?
For \alphaetc\ $=0$, $\Sigma\sim 1/p^{1-2}$
as discussed in section 3.2.
As \alphaetc\ is increased the shape of $\Sigma$ changes in a subtle
way at first by developing a tail for $p\approx M_{ETC}$
which enhances the condensate somewhat.
Then at some strong value of \alphaetc\ there is a dramatic
change in the solution, and $\Sigma$ becomes quite flat for larger
values of \alphaetc\  and the value of $\Sigma(0)$
begins to rise as quite a sensitive function of \alphaetc\ as it
approaches its NJL solution.

The above  effect has been observed in all the approaches mentioned
above but it is most simply studied in a theory
in which $\alpha_{TC}$ does not run and the effect of ETC boson
exchange is approximated by the four fermion approximation.
Apart from a Casimir this toy ETC theory is just quenched QED
which has been studied extensively by Bardeen {\it el al}
(1986, 1990b), Appelquist {\it et al} (1988c)
Kondo {\it et al } (1989), Dagotto {\it et al} (1990)
and Curtis and Pennington (1993).
For quenched QED one may plot a criticality curve as in figure 12
which shows the separation between the broken chiral symmetry phase
of the theory and the unbroken phase of the theory as a function
of the gauge coupling $\alpha$ and the four fermion coupling $\lambda$.
The dimensionless $\lambda$ is normalised so that for $\alpha=0$
chiral symmetry is broken for $\lambda>1$ (the NJL point).
In the pure gauge limit $\lambda=0$ chiral symmetry is broken for
$\alpha>\alpha^c = \pi/3$. On the criticality curve one has,
\beq
\lambda=\frac{1}{4} (1+\sqrt{1-\alpha / \alpha^c})^2
\eeq
The anomalous dimension increases along the  curve from $\gamma=1$
(recall that this is quenched QED so that the coupling does not run)
to $\gamma=2$ according to,
\beq
\gamma=1+\sqrt{1-3\alpha/\pi}
\eeq

In such theories a phenomenologically
acceptable top quark mass has a price associated with it.
Firstly it seems unlikely that one can account for the mass
ratio $m_t/m_b$ without overinfecting the $\rho$ parameter
(Appelquist {\it et al} 1989, Holdom 1989).
Secondly the required condensate enhancements can only be achieved
at the expense of some fine-tuning.

The fine-tuning of \alphaetc\ (or equivalently $\lambda$)
is in turn associated with the appearance of light scalar bound states
(Chivukula {\it et al} 1990).
As Miransky (1991) has emphasised light bound states are always
associated with finely tuned theories with $\gamma\approx 2$.
Miransky's argument can be phrased very simply in terms of the
physical dimension of the operator $\bar{\psi}\psi$
which approaches the dimension of a scalar field $\phi$
as $\gamma$ approaches 2,
\beq
D_{\bar{\psi}\psi}=3-\gamma
\begin{array}{c}
\gamma \rightarrow 2 \\
\longrightarrow
\end{array}
1=D_{\phi}
\eeq
The fine-tuning is asociated with how close to the critical line
one's parameters are chosen to be. The scalar fields appear
as heavy, broad resonances, depending on the parameters of the
theory ( Appelquist {\it et al} 1991).

This phenomenon is the basis for the top quark condensate model
to be discussed in section 5.

\section{Experimental Prospects}

\subsection{Longitudinal W and Z scattering at the LHC }

We now discuss the experimental signatures of TC, beginning in this
section with the minimal TC model.
As we saw earlier, the three would-be GB's of this model are
eaten by the $W,Z$ and become their $L$ components.
There is no Higgs boson in this model, but there is
a broad scalar resonance at the TeV scale,
which is the TC analogue of the Higgs boson.
In fact the entire techni-hadron spectrum at the TeV scale
is just a scaled-up version of the ordinary QCD hadron spectrum
as indicated in table 2. The technimesons have the same quantum numbers
as the mesons of QCD but with isospin being replaced by techni-isospin.
We have simply scaled up the masses by a factor of 2650,
assuming a QCD-like TC theory based on $SU(3)_{TC}$.
Note that in addition to these
technimesons, there will also be TeV scale technibaryons.
The lightest technibaryon may be absolutely stable
and is a candidate for the dark matter of the Universe
(Nussinov 1985).

\begin{table}
\caption{Table 2.
Comparison of meson and technimeson spectrum in
minimal TC and QCD with one quark doublet.}
\begin{tabular}{|c|c|c|c|c|} \hline\hline
meson/technimeson & $J^{PC}$ & $I^G$ & meson mass & technimeson mass \\ \hline
$\pi^{\pm}/\pi^{\pm}_{TC}$ & $0^-$ & $1^-$ & 139.6 MeV & eaten \\ \hline
$\pi^{0}/\pi^{0}_{TC}$ & $0^{-+}$ & $1^-$ & 135.0 MeV & eaten \\ \hline
$\eta^{0}/\eta^{0}_{TC}$ & $0^{-+}$ & $0^+$ & 549 MeV & 1.5 TeV \\ \hline
$f^{0}/f^{0}_{TC}$ & $0^{++}$ & $0^+$ & 1400 MeV & 3.7 TeV \\ \hline
$\rho^{\pm}/\rho^{0}_{TC}$ & $1^{-}$ & $1^+$ & 770 MeV & 2.0 TeV \\ \hline
$\rho^{0}/\rho^{0}_{TC}$ & $1^{--}$ & $1^+$ & 770 MeV & 2.0 TeV \\ \hline
$\omega^{0}/\omega^{0}_{TC}$ & $1^{--}$ & $0^-$ & 782 MeV & 2.1 TeV
\\ \hline \hline
\end{tabular}
\end{table}

If minimal TC is correct then there will be no light Higgs boson,
but instead a ``mini-desert'' up to the TeV scale where all the
technimesons are. To test the minimal TC model we therefore must
wait for LHC energies at which we can hope to find evidence for
some of the technimesons in table 2.
In fact the best way to study minimal TC is to study longitudinal
$W$ and $Z$ scattering processes such as:
\beqarr
W_L W_L & \rightarrow & W_L W_L \nonumber \\
W_L Z_L & \rightarrow & W_L Z_L
\eeqarr
Since the L components of the gauge bosons correspond
to technipions, this amounts to technipion-technipion scattering.
More precisely, there is an equivalence theorem
proved to one-loop order and discussed by
Chanowitz and Gaillard (1985) (based on earlier work
as discussed by these authors).
According to the equivalence theorem,
in the ultra-relativistic limit
the longitudinal gauge boson scattering amplitudes
calculated in unitary gauge are equal to the corresponding
GB scattering amplitudes calculated in renormalisable gauge.
At finite energies $E_i$, there are corrections as shown below
\beq
M(W_L(p_1)W_L(p_2)\ldots ) = M(\pi (p_1) \pi (p_2)\ldots ) + 0(M_W/E_i)^2
\eeq

Armed with the equivalence theorem,
low energy pion-pion scattering results
can be applied to technipion-technipion scattering
and hence to longitudinal gauge boson scattering.
The most convenient and systematic way of discussing this is to use
the chiral lagrangian (see Georgi 1984).
For example to lowest order the chiral lagrangian yields,
\beq
M(W_L^+W_L^-\rightarrow Z_LZ_L)=\frac{s}{F_{\pi}^2}
\eeq
which leads to a cross-section which rises with
the energy squared $s$
and eventually violates partial wave unitarity
for $\sqrt{s}\approx  1.7$ TeV.
The above result has been extended to order
$0(s^2/F_{\pi}^4)$ by several authors, for example
Dawson and Willenbrock (1989).
In the minimal standard model,
the amplitude is cut off by s-channel Higgs exchange,
which cancels the bad high energy (unless the Higgs
mass is very large).
In TC the amplitude is cut off by the
s-channel technimeson resonances,
but since these masses are very large the scattering amplitude
must necessarily be quite strong.

There has been an enormous amount of work on trying to determine
exactly how the scattering amplitude is cut off in TC theories
and consequently what the precise signature of TC would be at the
LHC. The big question is how to distinguish the TC signature
from a very heavy Higgs boson in the standard model.
The issues are quite complicated and rather than discuss them
here we refer the reader to two examples of recent literature.
The first example is the so called BESS model of
Casalbuoni {\it et al} (1987, 1991) which assumes
a triplet of new vector resonances (the technirhos).
The second example is the chiral lagrangian approach of
Dobado {\it et al} (1991).
As discussed by Pauss (1990)
LHC studies indicate that providing
the (charged) technirho is not much heavier than about 1.5 TeV
the WZ channel would, after suitable cuts,
exhibit a peak above the
background in both the BESS and the DHT approaches,
assuming an LHC of $\sqrt{s}=16$ TeV, with
an integrated luminosity of $10^5\ pb^{-1}$,
as shown in figure 13.

\subsection{Pseudo-Goldstone bosons from a single techni-family}

In this section we consider the single techni-family scenario,
and briefly discuss the resulting pseudo-Goldstone boson phenomenology.
A more detailed discussion can be found in Farhi and Susskind
(1981) who were the first to suggest such a model.
Consider a TC model based on the gauge group,
\beq
SU(N)_{TC}\otimes SU(3)_C \otimes SU(2)_L \otimes U(1)_Y
\eeq
and with a single techni-family,
\beq
\begin{array}{ccl}
{Q_L^{TC}}^{\alpha} =
\left( \begin{array}{c}U_L^{TC} \\ D_L^{TC} \end{array}
\right)^{\alpha} &\sim& (N,3,2,1/6)\\
{U_R^{TC}}^{\alpha} &\sim& (N,3,1,2/3)\\
{D_R^{TC}}^{\alpha} &\sim& (N,3,1,-1/3)\\
{L_L^{TC}}^{\alpha} =
\left( \begin{array}{c}N_{L}^{TC} \\ E_L^{TC} \end{array}
\right)^{\alpha} &\sim& (N,1,2,-1/2)\\
{E_R^{TC}}^{\alpha} &\sim& (N,1,1,1)\\
{N_R^{TC}}^{\alpha} &\sim& (N,1,1,0)
\end{array}
\eeq
The technifermions which carry technicolour and QCD colour are referred
to as techniquarks, while the technifermions which carry technicolour
but not QCD colour are called technileptons. Note that the
right-handed technineutrino $N_R^{TC}$ is required by anomaly
cancellation, and cannot be given a Majorana mass without breaking
$SU(N)_{TC}$. If we had chosen the TC gauge group to be
$SO(N)_{TC}$ then the right-handed
technineutrino would not be required by anomaly
cancellation, and it could be given a Majorana mass without breaking
$SO(N)_{TC}$. This is simply because the $N$ representation of
$SU(N)$ ($SO(N)$) is complex (real).
The quarks and leptons transform as usual and are technicolour
singlets
\beq
\begin{array}{ccl}
{Q_L}^{i} = \left( \begin{array}{c}U_L \\ D_L \end{array}
\right)^{i} &\sim& (1,3,2,1/6)\\
{U_R}^{i} &\sim& (1,3,1,2/3)\\
{D_R}^{i} &\sim& (1,3,1,-1/3)\\
{L_L}^{i} = \left( \begin{array}{c}\nu_{L} \\ E_L \end{array}
\right)^{i} &\sim& (1,1,2,-1/2)\\
{E_R}^{i} &\sim& (1,1,1,1)\\
{\nu_R}^{i} &\sim& (1,1,1,0)
\end{array}
\eeq
where $\alpha=1 \ldots N$ is an $G_{TC}$ index,
and $i=1 \ldots 3$ labels the three families of quarks and leptons.

In the limit that QCD and electroweak interactions are switched
off, the TC sector of the model respects a large chiral symmetry
\beq
SU(8)_L\otimes SU(8)_R
\eeq
since there are eight technifermions arranged in the vectors,
\beq
(U^{red},U^{blue},U^{green},D^{red},D^{blue},D^{green},E,N)_{L,R}
\eeq
Electroweak symmetry is broken by the condensates,
\beq
<\bar{U}^{TC}_LU^{TC}_R>=
<\bar{D}^{TC}_LD^{TC}_R>=
<\bar{E}^{TC}_LE^{TC}_R>=
<\bar{N}^{TC}_LN^{TC}_R> \ne 0 \label{onefamilycondensates}
\eeq
Since the techniquark condensates have QCD colour,
there are really eight separate condensates above, which break
the chiral symmetry down to $SU(8)_{L+R}$. The electroweak symmetry
is now broken by the equivalent of four separate technidoublets
(red, blue, green doublets of techniquarks
and a doublet of technileptons)
and so the technipion decay constant required to give the correct
$W,Z$ masses is now,
\beq
F_{\pi}=\frac{246}{\sqrt{4}} GeV=123 GeV
\eeq
According to Goldstone's theorem we would expect $8^2-1=63$ massless
GB's, one for each broken generator. Three of these will get eaten by
the Higgs mechanism, leaving 60 GB's in the physical spectrum.

According to our previous discussion, we would not expect
these to be massless GB's but light PGB's due to the
explicit chiral symmetry breaking effects which we have so far
ignored. The PGB mass spectrum was estimated by Peskin (1980)
and Preskill (1981). These authors also studied the whole
question of vacuum alignment, i.e. the problem of which condensates
form and which symmetries are broken in TC theories.
For example it is possible in principle for a techniquark to condense
with a technilepton, thereby breaking QCD gauge symmetry. Clearly this
would be undesirable, and as it turns out is disfavoured
by QCD gluon exchange which acts as a small perturbation which
breaks the degeneracy of the technifermion condensates.
The condensates which respect QCD correspond to a slighly lower vacuum
energy than the condensates which break QCD, and since the condensates
are otherwise degenerate this is enough to tip the balance in favour
of QCD-conserving condensates. The condensates which we have assumed
in equation \ref{onefamilycondensates} are in fact the ones which are
prefered by these kinds of arguments. The same formalism which is used
to study the question of vacuum alignment also yields the estimates
of PGB masses shown in figure 14 taken from Peskin (1980).
As mentioned, these mass estimates are increased in walking
TC theories since condensate enhancement also enhances PGB masses.

In figure 15 the PGB spectrum is shown for an orthogonal
TC gauge group $SO(N)_{TC}$ in which the symmetry breaking
for a single technifamily is $SU(16)\rightarrow SO(16)$.
An example of an orthogonal TC model was given in section 2.3.
There are $(16^2-1)-(16\times 15)/2=135$ broken generators,
leading to 132 PGB's (plus 3 eaten would-be GB's).

Note that in both figures 14 and 15
the class 1 PGB's corresponding to the colour singlet states
receive mass from electroweak and ETC perturbations and so are
not expected to be massless. Indeed Eichten and Lane (1980) estimate
such PGB's to have masses up to $40$ GeV from ETC effects.
Due to condensate enhancement, much larger masses
are also possible.

The 60 PGB's of the $SU(N)_{TC}$ model
consist of the following states,
\beqarr
\bar{Q}\gamma_5\lambda^{\alpha}\tau^aQ \nonumber \\
\bar{Q}\gamma_5\lambda^{\alpha}Q \nonumber \\
\bar{Q_i}\gamma_5\tau^aL \nonumber \\
\bar{Q_i}\gamma_5L \nonumber \\
\bar{L}\gamma_5\tau^aQ_i \nonumber \\
\bar{L}\gamma_5Q_i \nonumber \\
\bar{Q}\gamma_5\tau^aQ + \bar{L}\gamma_5\tau^aL \nonumber \\
\bar{Q}\gamma_5\tau^{\pm}Q - 3\bar{L}\gamma_5\tau^{\pm}L \nonumber \\
\bar{Q}\gamma_5\tau^{3}Q - 3\bar{L}\gamma_5\tau^{3}L \nonumber \\
\bar{Q}\gamma_5Q - 3\bar{L}\gamma_5L
\eeqarr
where $Q=(U,D),\ L=(N,E)$,
$\lambda^{\alpha}$ ($\alpha=1 \ldots 8$) are the Gell-Mann
colour matrices and $\tau^a$ ($a=1 \ldots 3$) are the Pauli
isospin matrices.

The experimental signatures of the PGB's were studied by Ellis {\it et al}
(1981), Dimopoulos (1980), and many other authors at around this time.
A more recent discussion can be found in Eichten {\it et al} )(1986).
We divide our brief discussion into searches for colour octet PGB's,
colour triplet PGB's (i.e. leptoquarks) and colour singlet
PGB's (which resemble charged Higgses).

As it turns out, it is relatively straightforward to find the colour octet
PGB's at the LHC, but much harder (but not impossible) at the Tevatron.
Consider the colour octet neutral state $P_8^0=(U\bar{U}+D\bar{D})_8
/\sqrt{2}$. This is a techni-isospin singlet and can be produced
singly in hadronic collisions with a cross-section
$\frac{d\sigma}{dy}\approx 1 (10^{-2})$nb at the LHC (Tevatron)
(for rapidity $y=0$).
The $P_8^0$ can decay back into $gg$ or into $t\bar{t}$ if
kinematically allowed. The first signal at the Tevatron may be
an enhancement of the top quark
production cross-section, as discussed by Eichten {\it et al} (1986)
and Appelquist and Triantaphyllou (1992b).
In fact the CDF cross-section for $t\bar{t}$ production
does appear to be somewhat higher than
standard model expectations (Abe {\it et al} 1994).

The colour triplets are examples of leptoquarks.
There are many of them in the spectrum,
consisting of colour triplet combinations such as
$U\bar{N}$, $U\bar{E}$, $D\bar{N}$, $D\bar{E}$ and their antiparticles.
At the LHC or HERA they
are copiously pair produced and tend to decay
into heavy quarks and leptons with relatively background-free signatures.
For example a typical signature of a leptoquark pair might be
$t\bar{t}\tau \bar{\tau}$ which has a particularly low background.

The colour singlets are similar to charged and neutral Higgs bosons.
The best place to look for them is the clean environment provided
by the LEP collider, although they should also be seen at the LHC.
The neutral PGB's do not have a tree-level coupling
to the $Z$ boson, however, which should enable it to be
distinguished from neutral Higgs bosons.
They couple to gauge bosons via the triangle anomaly,
with technifermions running round the loop,
as discussed in some detail by Ellis {\it et al} (1981).
The charged PGB's couple to the photon and $Z$ by tree-level
couplings which resemble those for charged Higgs,
and like charged Higgs tend to decay into the heaviest fermions around.

Finally note that the coloured PGB's may re-scatter into
eaten technipions, and hence may enhance the rates of longitudinal
gauge boson scattering, as observed by Bagger, Dawson and Valencia
(1991).

\subsection{Precision Electroweak Measurements}

The subject of this section is very specialised and rapidly
changing, as new data becomes available as well as new theoretical
variables with which theory can be compared to experiment.
Yet this is also a subject of immense importance to TC theories,
because it has been widely reported that the recent data disfavour
TC theories, a claim that has been disputed by several authors.

The discussion of radiative electroweak corrections
is commonly resticted to the oblique corrections
(Kennedy and Lynn 1989)
(i.e. the vacuum polarisation graphs only) which can be
very conveniently parametrised in terms of three parameters
S, T and U (Peskin and Takeuchi 1990, 1992).
One can argue that present data do not restrict the U parameter
very much which can thus be ignored.

The S parameter is defined by,
\beq
S\equiv 16\pi \frac{d}{d q^2}
[\Pi_{33}^{new\ physics}(q^2)- \Pi_{3Q}^{new\ physics}(q^2)]|_{q^2=0}
\eeq
The T parameter is defined by,
\beq
\alpha T \equiv \frac{e^2}{\sin^2\theta_WM_W^2}
[\Pi_{11}^{new\ physics}(0)- \Pi_{33}^{new\ physics}(0)]
\eeq
where $\Pi_{ij}^{new\ physics}$ are the gauge boson self-energies,
with indices $i,j=1,3$ refering to $SU(2)_L$ currents
and $j=Q$ refering to the electromagnetic current,
$M_W$ is the $W$ mass, $e$ is the electromagnetic charge,
and $q$ is the momentum flowing through the gauge boson propagator.
In writing the self-energies we have assumed that the total
self-energy can be divided into two parts,
\beq
\Pi_{ij}(q^2)=\Pi_{ij}^{standard\ model}(q^2)+\Pi_{ij}^{new\ physics}(q^2)
\eeq
and so the quantities $S,T$ are only concerned with the new physics
contributions, relative to some definition of the standard model
(i.e. some top and Higgs mass must be assumed).
In order for these definitions
to be useful, it is necessary to assume that
the new physics consists of new particles whose mass is very much
greater than the $W,Z$ masses. The point is that the physical
measured quantities such as the $Z$ mass and width,
the electroweak coupling constants, the $W$ mass,
the various forward-backward asymmetries, and so on
are measured at $q^2=M_Z^2$ or $q^2=M_W^2$ or $q^2=0$
and so involve quantities like $\Pi_{ij}^{new\ physics}(M_Z^2)$,
$\frac{d}{d q^2} \Pi_{ij}^{new\ physics}(q^2)|_{q^2=M_Z^2}$,
and similar expressions with $M_Z\rightarrow M_W$.
Since these values of $q^2$ are assumed small compared to
the masses of the new particles running round the loops,
it is assumed that
\beq
\Pi_{ij}^{new\ physics}(M_Z^2)\approx \Pi_{ij}^{new\ physics}(0)
+ M_Z^2\frac{d}{d q^2} \Pi_{ij}^{new\ physics}(q^2)|_{q^2=0} \label{approx1}
\eeq
\beq
\frac{d}{d q^2} \Pi_{ij}^{new\ physics}(q^2)|_{q^2=M_Z^2}
\approx
\frac{\Pi_{ij}^{new\ physics}(M_Z^2)- \Pi_{ij}^{new\ physics}(0)}
{M_Z^2} \label{approx2}
\eeq
and similar expressions with $M_Z\rightarrow M_W$.
These approximations correspond to the assumption that
the self-energy functions have a linear slope over the
small (compared to new physics masses) $q^2$ region
$0-M_{W,Z}^2$. With this approximation, all the physical
quantities of interest may be re-parametrised in terms
of $S,T$ (and $U$ which we have ignored).

The T parameter is essentially just the contribution to $\delta \rho$
from new physics, $\alpha T \approx \delta \rho^{new}$.
Crudely, T is a measure of the isospin splitting of the new
fermion doublets, while S is a measure of the number of new fermion
doublets. Detailed fits of all the electroweak data lead to
ellipses in the
S,T plane representing the allowed contribution from new
physics at, say 90\% c.l., with the origin S=T=0 corresponding
to the standard model at some reference value of $m_t,m_H$.
These fits tend to favour negative values of S and T such as
$S\approx -1,T\approx -0.5$, with positive values such as
$S\approx 1, T\approx 1$ disfavoured by the data.
The question is what does TC predict
for S and T? The answer is far from certain.
The problem is that S and T are difficult to estimate non-perturbatively.
Various authors have attempted to do so however using various
techniques ( Peskin and Takeuchi 1990, 1992,
Cahn and Suzuki 1991, Golden and Randall 1990,
Holdom and Terning 1990, Appelquist and Triantaphyllou 1992a,
Sundrum and Hsu 1991, Chivukula {\it et al} 1992a).
It turns out that T (or equivalently $\delta \rho$)
is highly model dependent.

The S parameter is of more interest due to its relative
model independence.
On the other hand S is very difficult to estimate in strongly
coupled theories such as TC. Peskin and Takeuchi (1990, 1992)
estimated S in TC from a scaled-up QCD dispersion relation
(assuming techni-isospin conservation and parity) and
concluded that $S\approx 1.6$ for a technifamily and
$S\approx 0.5$ for a single doublet (assuming $N=4$ in both cases).
The perturbative result which essentially counts the number
of electroweak doublets is $S\approx NN_D/(6\pi)$ where
$N$ is the number of TC's and $N_D$ is the number of doublets
(e.g. $N_D=4$ for a technifamily).
The QCD estimates are about a factor of 2 larger than the
perturbative estimates would yield.
Indeed several authors have pointed out that in walking TC
or strong ETC where $\Sigma$ is flattened then the estimates for
S and T should resemble more closely the perturbative estimates
(see for example Appelquist and Triantaphyllou 1992a).
But even in the perturbative limit the $SU(4)_{TC}$ technifamily
will still contribute $S=0.85$ which is on the fringes of the present
experimentally allowed range. The only conclusion from this
is that the techni-family is being severely challenged by
this data. I would go even further and say that unless
there are any get-outs, the techni-family is effectively
ruled out by this data. However as we shall see there are
get-outs available, so the situation is not
entirely clear cut.

I wish to emphasise once again that estimates of S are made
using a variety of approaches, which are not all mutually consistent.
In the midst of all the confusion, it seems to me that
one approach is particularly elegant, namely the quasi-perturbative
methods developed by Holdom {\it et al} (1989,1990)
and Terning (1991).
These methods draw their inspiration
from the Pagels-Stokar approach which calculates
pion decay constant using an integral sum rule in which the only input
is $\Sigma(p^2)$ (and its first derivative).
In the presence of isospin breaking one may extend the Pagels-Stokar
result in order to calculate $F_{{\pi}^{\pm}}$ and
$F_{{\pi}_3}$ from $\Sigma_U \neq \Sigma_D$. One then computes,
\beq
\alpha T=\delta \rho ^{new}
=M_{W^{\pm}}^2/M_{W_3}^2 - 1
=F_{{\pi}^{\pm}}^2/F_{{\pi}_3}^2-1
\eeq
In a similar spirit one may derive a formula for S (valid in
the isospin limit) as an integral sum rule in terms of
$\Sigma(p^2)$ and its higher derivatives.
In these approaches the gauge boson loop
corrections are approximated by loops of
technifermions with a running self-mass function $\Sigma (p^2)$,
which may be estimated from a gap equation.
Such estimates effectively ``mock-up'' the corrections
from heavy technirho exchange as well as the other heavy vector
resonances. They do not include contributions from
light PGB's which must be included separately and will therefore
give additional contributions ( Peskin and Renken 1983,
Golden and Randall 1991,
Holdom and Terning 1990). Luty and Sundrum (1993) have even
suggested that under some circumstances the PGB contribution
to S could turn out to be negative.

Several authors have considered sources of new physics which can yield
negative contributions to $S,T$. One example is
the observation that Majorana neutrino masses can lead
to negative $S,T$ (Gates and Terning 1991). As observed by
Gates and Terning, a Majorana mass for the technineutrino
would also be expected to lead to similar effects.
Using the non-local effective theory
methods, and assuming an
orthogonal $SO(N)_{TC}$ gauge group,
one can obtain a negative contribution
to S and T by giving the techni-neutrino a Majorana mass
(Evans {\it et al} 1993b, 1994a). In $SU(N)_{TC}$ a technineutrino mass
is of course forbidden by the gauge symmetry.
This may help with models such as the chiral ETC model
(King 1989b, King and Mannan 1992) which are
based on $SO(7)_{TC}$ and so
allow a right-handed technineutrino mass term.

So far we have restricted our attention to oblique
radiative corrections, i.e. corrections
to gauge boson propagators coming from loops of
heavy new particles. What about non-oblique corrections,
coming from vertex corrections for example?
In general, it turns out that vertex and box diagrams
are suppressed relative to the oblique corrections,
essentially due to small Yukawa couplings which enter in these
diagrams. However in certain cases the vertex corrections
can be important for TC models, the prime example
being the $Zb\bar{b}$ vertex. There are two sources
of corrections to this vertex in TC models,
namely from ETC gauge boson exchange (Chivukula {\it et al} 1992b,
1994)
and from PGB exchange (Xiao {\it et al} 1994).
The ETC gauge boson exchange contribution
is shown in figure 16 in the case where the ETC gauge group
commutes with the standard model gauge group,
so that the ETC gauge boson is colour and electrically neutral.
Such diagrams alter the left-handed gauge coupling
$g_L$ by an amount,
\beq
\delta g_L \sim \frac{m_t}{16\pi F_{\pi}}
\frac{e}{\sin \theta_W \cos \theta_W}
\eeq
where we have assumed that the top mass is given by
\beq
m_t \approx 4 \pi \frac{F_{\pi}^3}{M_{ETC}^2}
\eeq
The experimentally relevant quantity is the ratio,
\beq
\Delta_R=\frac{\delta (\Gamma_b/\Gamma_{hadrons})}
{\Gamma_b/\Gamma_{hadrons}}
\eeq
(where hadrons do not include $b$ quarks)
which LEP can measure to a precision of at least $2\%$.
The standard model radiative corrections to $\Delta_R$ are
about $-2\%$. The ETC corrections are
\beq
\delta \Delta_R \sim -3.7 \%\left( \frac{m_t}{100 \ GeV} \right)
\eeq
Thus this ETC model is ruled out since the top quark is
much heavier than 100 GeV.
PGB contributions to $\Delta_R$ can range from
0 to -10\% depending on the exact PGB spectrum (Xiao {\it et al} 1994).
As usual there are ways out. For example, Evans (1994b)
points out that if the ETC scale is boosted by strong ETC
effects then the constraint may be avoided. This requires
fine tuning of several per cent, however. Alternatively,
one may suppose that the top quark does not gain
its mass from ETC gauge boson exchange, but instead
gains its mass from the top quark condensate.
Once again this motivates the low scale TC scenario in the
next section.

The above discussion has been based on the $S,T,U$ parameters.
It is only fair to point out that there is an alternative
set of parameters in the literature called $\epsilon_1,
\epsilon_2, \epsilon_3$ and $\epsilon_b$,
introduced by Altarelli et al (1992, 1993).
The first three of these parameters play the role of $S,T,U$
but are more model-independant and physical. To be precise,
they are directly related to the physically measured quantities
$M_W/M_Z$, $\Gamma_l$ and $A_{FB}^l$, which are all measured
on the mass shell of the gauge boson. The last of the variable,
$\epsilon_b$ is related to $\Gamma_b$. The $\epsilon$'s have
the nice property that they are zero in the (QED and QCD improved)
Born approximation. Current values of these parameters show that they
are different from zero at the 2$\sigma$ level, and are nicely consistent
with a large top mass $m_t\approx 170$ GeV. Are technicolour theories
excluded by these constraints? The answer cannot be yes, because
with unconventional dynamics such as present in Walking TC,
we simply cannot calculate or even reliably estimate the values
of the radiative corrections, as recently emphasised by Lane (1994).

\subsection{Low scale technicolour}

So far we have assumed that the TC confinement scale is
${\Lambda}_{TC} \sim 500$ GeV so that in order
to probe the TC dynamics one must wait for the LHC.
In this section we discuss the phenomenology of an
$SU(2)_{TC}$ technicolour model proposed by King (1993)
with a low technicolour confinement scale ${\Lambda}_{TC} \sim 50-100$ GeV.
Such a low technicolour scale may give rise to
the first hints of technicolour being seen at LEPI and
spectacular technicolour signals at LEPII and the Tevatron.

How do we achieve a low TC scale and still break
electroweak symmetry strongly enough?
The answer is that we use the idea of strong ETC as discussed
in section 3.3 to enhance the TC condensate sufficiently
to give the correct $W$ mass. In other words we rely on some
four-technifermion operators which help to break electroweak symmetry,
so that TC does not have to do all the work itself,
thus allowing the TC scale to be reduced.
There may be dominant technifermions which are subject to the
strong operators and sub-dominant technifermions which are not.
The sub-dominant technifermions
will condense at a TC low scale
leading to the striking experimental signatures advertised above.
There is one other feature of the model worthy of note.
We shall assume that ordinary fermions (quarks and leptons) do
not acquire masses from the ETC mechanism, but instead generate
their own masses from heavy quark and lepton condensates
(e.g. the top quark condensate discussed in section 5).
Our motivation for this is the desire to account for the top mass without
running into the phenomenological difficulties that ETC faces.

Consider the gauge group
\beq
SU(2)_{TC}\otimes SU(3)_C\otimes SU(2)_L\otimes U(1)_Y
\eeq
where we have added to the standard model gauge group
a new confining QCD-like gauge group $SU(2)_{TC}$ which
is asymptotically-free and confines at ${\Lambda}_{TC}\sim 50-100 GeV$.
The technifermions in this model may be written,
\beq
\begin{array}{ccl}T_L=\left( \begin{array}{c}P_L \\ M_L \end{array}
\right) &\sim& (2,1,2,0)\\
{P_R}   &\sim& (2,1,1,1/2)\\
{M_R} &\sim& (2,1,1,-1/2)
\end{array}
\eeq
\beq
\begin{array}{ccl}t_L^i=\left( \begin{array}{c}p_L^i \\ m_L^i \end{array}
\right) &\sim& (2,1,2,0)\\
{p_R^i}   &\sim& (2,1,1,1/2)\\
{m_R^i} &\sim& (2,1,1,-1/2)
\end{array}
\eeq
where we have denoted the dominant technifermions by upper case letters,
and sub-dominant technifermions by lower case letters labelled by
$i=1,\ldots n_F-1$ where $n_F$ is the number of lepton families.
Henceforth we shall assume for simplicity that $n_F=3$.

To begin with let us consider the dominant
technidoublet $T=(P,M)$.
We assume operators of the form
$G_T(\bar{T}_{L}T_{R})(\bar{T}_{R}T_{L})$.
These operators are associated with a dimensional
scale $\Lambda \sim 1$ TeV.
For example in section 5.5 we shall discuss
a model in which the four-fermion operators will
arise from heavy (1 TeV) gauge boson exchange.
There are similar contact terms
of the form $G_{\tau}(\bar{\tau}_{L}\tau_{R})(\bar{\tau}_{R}\tau_{L})$
which induce a
tau lepton condensate.
Since we assume a tau condensate then $G_T\approx G_{\tau}$
must therefore be strong, leading to a condensate
$<\bar{P}P+\bar{M}M>\neq 0$.
If the operator respects isospin then the
pattern of symmetry breaking
expected is just
\beq
SU(2)_L\otimes SU(2)_R\rightarrow SU(2)_{L+R} \label{chiral}
\eeq
yielding a triplet of technipions
$\Pi_{TC}^{\pm ,0}\sim \bar{T}\sigma^{\pm ,3}\gamma_5T$,
where $\sigma^a$ are the Pauli matrices.
In such a theory the technipion decay constant may be much larger
than $\Lambda_{TC}$ since chiral symmetry breaking is driven
mainly by the above contact operator.
We assume $F_{TC}\sim 245 GeV$.
The technipions associated with the dominant
technidoublet get eaten,
and the remaining technihadrons have
masses set by the enhanced dynamical masses of the technifermions,
and are in the LHC range.

Now let us extend our discussion to include one of the sub-dominant
technidoublets in this model.
There are now two technidoublets $T=(P,M),\ t=(p,m)$, which have identical
quantum numbers.
The set of operators in this case are
of the form
\beqarr
G_T(\bar{T}_{L}T_{R})(\bar{T}_{R}T_{L}),\
G_t(\bar{t}_{L}t_{R})(\bar{t}_{R}t_{L}),\
G_{tT}(\bar{T}_{L}T_{R})(\bar{t}_{R}t_{L}) \label{Tt}
\eeqarr
where $G_t,G_{tT}\ll G_T$.
In this case the theory naturally splits into two parts,
a high scale TC sector associated with the technifermions $T$
which form condensates driven by the operators
associated with the scale $\Lambda \sim 1 TeV$,
as discussed above,
and a low scale TC sector associated with technifermions $t$
which form condensates driven by TC interactions at the scale
$\Lambda_{TC}\sim 50-100$ GeV.
The low scale TC sector
again has an approximate global symmetry as before,
but now there is a vacuum alignment problem
which depends on the relative strength of the contact term
and the TC gauge forces. The TC gauge forces tend to favour the
chiraly invariant condensate
$<\bar{t^c_L}t_L+\bar{t^c_R}t_R>\neq 0$,
while the contact terms prefer the chiral symmetry
breaking condensates,
$<\bar{t_L}t_R+\bar{t_R}t_L>\neq 0$.
We shall assume the latter condensates form
yielding a triplet of technipions
$\pi_{TC}^{\pm ,0}\sim \bar{t}\sigma^{\pm ,3}\gamma_5t$,
associated with a much smaller pion decay constant
$f_{TC}\sim 10-50 GeV$. These pions do not get eaten,
and remain in the physical spectrum.

The physical technipions $\pi_{TC}^{\pm,0}$ will receive a mass
from the mixed operator with coefficient $G_{tT}$.
The ``explicit'' masses of the
low scale technifermions resulting from this operator
are obviously just $m_{p,m}=G_{tT}<(\bar{T}_{L}T_{R})>$.
These masses break the chiral symmetry of the second
technidoublet resulting in a physical technipion mass
analogous to the way in which the physical pion mass
results from explicit quark masses.
The technipion mass $m_{\pi_{TC}}$ may be estimated by scaling up the
usual result for the ordinary pion mass $m_{\pi}$,
\beq
m_{\pi_{TC}}= m_{\pi}\sqrt{\left(\frac{m_p + m_m}{m_u + m_d}
\right)\frac{f_{TC}}{f_{\pi}}}.
\eeq
Using these equations
we find rather heavy technipion masses of order tens of GeV,
or perhaps hundreds of GeV, depending on $f_{TC}$.

The charged technipions decay via ordinary $W$ exchange
into the heaviest fermion channels available such as cb.
The physical neutral technipion
$\pi_{TC}^0$ will decay into two photons via a chiral symmetry
suppressed anomalous $\pi_{TC}^0\gamma \gamma$ coupling.
The partial width is given by (King 1993),
\beq
\Gamma (\pi_{TC}^0 \rightarrow \gamma \gamma)=
A_{\pi_{TC}^0\gamma \gamma}^2
\left( \frac{\alpha^2}{16\pi^3f_{TC}^2}\right)m_{\pi_{TC}^0}^3
\eeq
where in the present model $A_{\pi_{TC}^0\gamma \gamma}=
-(1/8)\left(\frac{m_{\pi_{TC}^0}}{\Lambda_{TC}}\right)^2$.
Despite its small partial width, this is expected to be an important
decay mode of the $\pi_{TC}^0$ leading to a striking signature.
For example, in 1993 the L3 collaboration at LEP
reported four events, one $e^+e^-\gamma \gamma$
and three $\mu^+\mu^-\gamma \gamma$,
each with two energetic photons with an invariant mass
$M_{\gamma \gamma}\approx 60 GeV$. This would be a typical
signature for a neutral technipion of mass 60 GeV.
Although in conventional models the
production rate would be very small (Lubicz 1993), in the present
model the technipion may be produced in association
with a virtual techniomega
which serves to enhance its production rate (King 1993).

Apart from the low-scale technipions, the technidoublet $t$
will give rise to technivector mesons $V$
analogous to the QCD vector resonances.
However here the masses of such technivectors will be
{\em an order of magnitude} smaller than table 2.
For example we may expect a $J^{PC}=1^{--}$
technirho triplet ${\rho}_{TC}^{\pm ,0}$ and techniomega singlet
${\omega}_{TC}$ with masses in the LEPII range 100-200 GeV.
Later on we shall consider extending this mass range to 200-300 GeV,
which is relevant for the Tevatron.
The vector masses may be estimated by scaling up
the ordinary $\rho$ and $\omega$ mass
$m_{\rho_{TC} ,\omega_{TC}}\approx m_{\rho ,\omega}\frac{f_{TC}}{f_{\pi}}$.
The ${\rho}_{TC}^0$ may be detected at LEPII via its
couplings to the photon and Z,
\beq
\frac{m_{{\rho}_{TC}}^2}{g_{{\rho}_{TC}}}
{\rho_{TC}}^{0,\mu}\left[eA_{\mu} +
\frac{e}{\tan 2\theta_w}Z_{\mu}\right] \label{Zrho}
\eeq
leading to resonances in R.
The techniomega $\omega_{TC}^0$ being associated with the
isosinglet current $\bar{t}\gamma^{\mu}t$ does not couple to the photon or Z,
but may have a direct coupling to leptons.

There are other models
in the literature which also have a low TC scale, for example
Eichten and Lane (1989). These authors considered the Tevatron
signatures of a light technirho.
The technirhos are produced via their electroweak couplings
to the $W$ and $Z,\gamma$.
The dominant decay of the technirho $\rho_{TC}$ is into
$\pi_{TC} \pi_{TC}$, but this channel is
not kinematically accessible if the technipions are heavier
than half the technirho mass. In this case the technirho will
decay dominantly into $\pi_{TC} W_L$, or $\pi_{TC} Z_L$, at a rate
suppressed by $(f_{TC}/F_{TC})^2$. For example we may have,
\beq
\bar{q}q\rightarrow Z^{\ast} \rightarrow \rho_{TC}^0
\rightarrow \pi_{TC}^+W^-
\eeq
with the technipion decaying into $tb$ or $cb$, leading to a
Tevatron signal consisting of
a $W$ recoiling against a di-jet, with the whole system
having an invariant mass equal to that of the technirho.

\section{Top Quark Condensates}

\subsection{Four-fermion theory}

We have seen that obtaining a large top quark
mass in TC theories is non-trivial. One is led to consider
strong ETC which as discussed in the section 3.3
involves the consideration
of four-technifermion operators which contribute to the gap equation.
Such theories are characterised by fine-tuning and light scalar
bound states. It is a small step from these types of theory to
top quark condensate theories,
first postulated by Nambu (1989),
and subsequently studied by Miransky {\it et al} (1989),
Marciano (1989) and
Bardeen {\it et al} (1990).
Shortly afterwards King and Mannan (1990)
showed how the effect of the four-fermion operator
could be replaced by heavy gauge boson exchange.
Two Higgs doublet models resulting from
top and bottom condensates were discussed by
Suzuki (1990a) and Luty (1990).
Much of the discussion of the present
section is taken from
Bardeen {\it et al} (1990).

The starting point of top quark condensate models
is to postulate a four-fermion operator,
\beq
G(\bar{Q}_Lt_R)(\bar{t}_RQ_L) \label{tttt}
\eeq
where $Q_L^T=(t_L,b_L)$ and we may write $G=(8\pi^2/3)\lambda /\Lambda^2$
where $\Lambda$ is called the ultraviolet cut-off, and is a dimensionful
parameter of the theory.
 From our experience with strong ETC we may regard the contact operator
as arising from some heavy gauge boson of mass $\Lambda$
being exchanged between top quarks. Such a heavy boson must couple
strongly to top quarks, as the coefficient $8\pi^2/3$ suggests.
In section 5.5 we shall discuss
a model in which such a heavy boson can emerge
but for now we shall live with the non-renormalisable operator
above.

The gap equation arising from the operator in equation \ref{tttt}
is represented by figure 18.
This yields a self-consistent equation for $m_t$,
\beq
m_t=6Gm_t\frac{i}{(2\pi)^4} \int^{\Lambda}
\frac{d^4k}{(k^2-m_t^2)}
=\Lambda^2(1-\frac{1}{\lambda})
\eeq
Assuming that $m_t\neq 0$ the top mass can be cancelled from both
sides of the equation, and the integration can be performed to yield,
\beq
m_t^2\ln \frac{\Lambda^2}{m_t^2}=\Lambda^2 - \frac{8\pi^2}{3G}
\eeq
For $\lambda \gg 1$ it is clear that $m_t=0(\Lambda)$.
For $\lambda<1$ the solution $m_t=0$ is prefered, and the gap equation
is trivial. If we fine-tune $\lambda= 1 + \epsilon$ then we can obtain
$m_t\ll \Lambda$.

We expect fine-tuning to be associated with the appearance
of light scalar bound states, similar to our discussion
in the section 3.3.
The light bound states may be revealed by looking for poles
in the $\bar{t}_Lt_R\rightarrow \bar{t}_Lt_R$ scattering
amplitude. A standard bubble summation can be performed,
which like the gap equation is valid in the large $N_c$ limit,
where $N_c=3$ is the number of colours.
The $1/N_c$ corrections have been studied by Hands {\it et al}
(1991).
The result of the bubble summation is to exhibit a pole in the
scalar $\bar{t}t$ channel (once the gap equation has been implemented
to cancel the leading divergence) at $p=2m_t$. This is the
Higgs boson pole. There are also poles in the pseudoscalar
channels $\bar{t}\gamma_5t$, $\bar{t}\gamma_5b$ and
$\bar{b}\gamma_5t$ at $p=0$ corresponding to the three
massless Goldstone bosons originating from the Higgs doublet.
Note that the mass of the Higgs boson and the top quark
are related, and are kept light by a common fine tuning.
The simple relation above indicates that the Higgs boson
is a bound state of $\bar{t}t$ with zero binding energy.
However, this result is only valid in the pure fermion approximation,
and it neglects any gauge corrections such as QCD gluon exchange.
It also neglects Higgs exchange corrections. The point is that the
Higgs boson must be regarded as a light propagating degree of freedom,
which can help to bind itself by a bootstrap mechanism
as originally envisaged by Nambu (1989).

\subsection{Renormalisation Group Approach}

There are two ways in which the gauge and Higgs corrections can
be calculated. The first way is to simply add a QCD gluon exchange
term to the right-hand side of the gap equation (King and Mannan 1990).
QCD gluon exchange corrections can also be added to the bubble
sums in a similar way (Chesterman {\it et al} 1991).
These are difficult calculations,
and do not include the important Higgs exchange corrections.
However there is a much easier method in which all corrections
(at leading order) may be conveniently considered, namely the
equivalent Lagrangian approach of Bardeen {\it et al} (1990).

The starting point of the equivalent Lagrangian approach is to
observe that the four-fermion operator in equation \ref{tttt}
may be expressed
in an alternative way in terms of an auxiliary complex scalar doublet
$H$ as follows,
\beq
(\bar{Q}_Lt_R)(\bar{t}_RQ_L)
=\bar{Q}_Lt_RH + h.c. - \frac{1}{G}H^{\dag}H \label{equiv}
\eeq
where $H=G(\bar{t}_RQ_L)$ will become the composite Higgs doublet.
The equivalent Lagrangian in equation \ref{equiv}
is valid at the energy scale
$\mu=\Lambda$. The effective low-energy Lagrangian at an energy
scale $\mu\ll \Lambda$ is obtained by integrating out the high-energy
fermion degrees of freedom
(Bardeen {\it et al} 1990).
It is straightforward to show that the low energy effective theory
at $\mu\ll \Lambda$ is just the minimal standard model, a few terms
of which are shown below,
\beq
L^{eff}=g_t\bar{Q}_Lt_RH + h.c
+|D_{\mu}H|^2 - M^2H^{\dag}H
- \frac{\lambda}{2}(H^{\dag}H)^2
+0(\frac{1}{\Lambda^2}) \label{eff}
\eeq
but with the compositeness boundary conditions,
\beq
g_t(\mu),
\lambda(\mu)\rightarrow \infty|_{\mu \rightarrow \Lambda}
\label{bc}
\eeq
These compositeness boundary conditions arise from the simple
requirement that at $\mu=\Lambda$ is satisfied.

Equations \ref{eff}, \ref{bc}
correspond to a quasi-fixed point of the standard
model (Pendleton and Ross 1981, Hill 1981, 1990),
and lead to predictions for the Higgs boson mass
and top quark mass as a function of the ultraviolet cut-off
$\Lambda$. One simply integrates the standard model
renormalisation group (RG) equations using the boundary conditions above.
This is the great power of this method, because the resulting
predictions have all the gauge corrections and Higgs
exchange corrections taken account of by the
standard model RG equations.
The words quasi-fixed point mean that the RG trajectory
of the Yukawa couplings $g_t(\mu)$ and $\lambda(\mu)$
tend to converge on a certain (fixed) low energy value
almost independently of their boudary value at $\mu=\Lambda$
providing that this boundary value exceeds a certain minimum
value.
It is this feature which is very important in giving reliable
predictions, since the precise boundary conditions in equation \ref{bc}
cannot be taken literally since they are infinitely far outside the
perturbative region.
The resulting predictions are shown in table 3 for three values
of cut-off $\Lambda$ and for both the top quark condensate model
and a degenerate four family model.

It must be said that the results are disappointing.
The top quark mass is always much heavier than the tentative
current experimental measurement of about 170 GeV.
One way to improve the situation is to make the top
quark condensate model supersymmetric (Clark {\it et al} 1990,
Carena {\it et al} 1992). This approach exploits the fact
that in the minimal supersymmetric standard model,
the triviality limit of the top quark mass is less
than in the standard model. There appears to be nothing wrong
with this approach, apart from the observation that the supersymmetric
gap equation leads to a critical coupling constant which is
very much larger than in the non-supersymmetric model.
However it is difficult to get excited about such models
because the scale of new physics is of order the Planck scale,
or grand unified scale. As discussed in section 5.4 it is quite
likely that such models with high scales are isomorphic to the
standard model ( or supersymmetric standard model).
What is much more interesting is if the scale of new physics
is reduced to the TeV region. This can be achieved in several ways.
One way is immediately apparent from table 3:
introduce a fourth family of fermions, so that the
top quark mass becomes unconstrained and the scale
of new physics may be reduced to the TeV region.
Note that the four-family case with $\Lambda=2$ TeV involves
no fine-tuning. With such a low scale of new physics,
one would expect experimental signatures from
longitudinal $W$ and $Z$ scattering experiments
at LHC. This was discussed by Chestermann and
King (1992).

\begin{table}
\caption{
Table 2. Predictions of top mass (t) and Higgs (h) mass as a function of
cut-off $\Lambda$. Similar predictions are also shown for
a degenerate fourth family ($m_{t'}=m_{b'}$).}
\begin{tabular}{|c|c|c|c|} \hline\hline
{$\bowtie$} & {$\Lambda=10^{17}GeV$}
& {$\Lambda=10 TeV$}   &       {$\Lambda=2 TeV$} \\ \hline
{$m_t$} & {220 GeV} & {450 GeV} & {-} \\
{$m_h$} & {250 GeV} & {600 GeV} & {-} \\ \hline
{$m_{t'}=m_{b'}$} & {200 GeV} & {400 GeV} & {$\approx 1 TeV$} \\
{$m_h$} & {240 GeV} & {550 GeV} & {$\approx 2 TeV$} \\ \hline
\end{tabular}
\end{table}

\subsection{A fourth family after all?}

One may wonder whether a fourth family is still a viable
possibility after the LEP measurement of the $Z$ width which
indicates three and only three light neutrinos?
If there is a fourth family then the fourth neutrino
must be heavier than half the $Z$ mass in order that
the $Z$ does not decay in this channel.
But having a fourth neutrino heavier than 45 GeV,
seems a little odd given that the first three neutrino
species are so light. However it turns out that
it is possible to write down models which quite naturally
lead to the desired neutrino hierarchy.
For example a singlet Majoron model
has been proposed by Hill {\it et al} (1990, 1991)
in which the four right-handed neutrinos
gain Majorana masses of order 1 TeV. This scheme then
predicts three species of Majorana neutrino near their laboratory mass
limits plus a heavy fourth neutrino.
The model also predicts massless Majoron
which is necessary in order that the heavier neutrinos
can decay into the lighter ones (via Majoron emission)
at a fast enough rate to
avoid over-closing the Universe.
However the Majorons also couple to the $Z$ via neutrino loops,
and since the $Z$ couples to all fermions,
the Majoron also may couple to any fermion.
As shown by Mohapatra and Zhang (1993) the see-saw four-family
model is in trouble because the Majorons associated with it
lead to conflicts with astrophysics. Essentially stars can
emit Majorons as a cooling mechanism, which constrains the
effective Majoron coupling to electrons, and this
constraint turns out to be violated in the case where the
symmetry breaking scale associated with the Majoron
is of order 1 TeV, as assumed
in the four-family see-saw model.

However there are other natural models of the
fourth family which are not subject to the above constraints.
The minimal four family model (King 1992c, 1992d)
introduces only one
right-handed neutrino (with zero Majorana mass),
rather then three.
This single right-handed neutrino has to be
shared amongst three left-handed neutrinos,
and results in a mass spectrum consisting of
three massless neutrinos plus a heavy Dirac neutrino.
Because the single right-handed neutrino is assumed to
be precisely massless total
lepton number $L$ is exactly conserved in this
model. However the separate lepton numbers $L_e,L_{\mu},L_{\tau}$
are not conserved. It is $L$ conservation which forbids
all Majorana neutrino masses and enforces the masslessness of
the first three neutrinos. A non-minimal four-family model
with a single right-handed neutrino was in fact proposed by
Babu {\it et al} (1988, 1989) before the LEP result of three
light neutrinos.
In the non-minimal four family model
the single right-handed neutrino is given a Majorana mass,
thereby breaking $L$ number and allowing the first three neutrinos
to acquire small masses from a two $W$ exchange mechanism.
The idea of the earlier non-minimal four-family model
was not to provide an explanation
for why the fourth neutrino was heavier than $M_Z/2$
but rather to discuss the
neutrino masses and mixings in such a model.
In the minimal four-family model of King (1992c),
which was invented independently,
the fact that lepton number is conserved leads to a particularly
simple parametrisation of lepton mixing as discussed by King (1992d).
In summary,
the minimal four-family model provides an elegant mechanism
for having three massless neutrinos plus one heavy Dirac neutrino,
and therefore the idea of a fourth family should certainly not be excluded
on the basis of the LEP measurement of the $Z$ width.
The above models are of course independent of the idea
of DEWSB.

\subsection{Irrelevant Operators}

We wanted to account for a heavy top quark, but the results
in table 2 look like overkill. The top quark mass of $220$ GeV
is just too heavy.
The superheavy scale $\Lambda=10^{17}GeV$ implies fine-tuning,
and no prospect of direct experimental verification of the theory.
However the predictions in table 2 have been criticised
by Suzuki (1990b) and
later by King and Mannan (1991)
due to the neglect of higher dimensional or ``irrelevant'' operators,
which if they have sufficiently large dimensionless coefficients,
may alter the predictions significantly.
A generalised NJL model has been proposed by Hasenfratz {\it et al} (1991),
\beq
{L}_{NJL}^{gen}=
G(\bar{t}t+\frac{\chi}{\Lambda^2}\partial_{\mu}\bar{t}\partial^{\mu}t)
(1+\frac{f_1}{\Lambda^2}\partial^2+\ldots)
(\bar{t}t+\frac{\chi}{\Lambda^2}\partial_{\mu}\bar{t}\partial^{\mu}t)
+ \ldots \label{gen}
\eeq
where the term with coefficient $f_1$ represents the first term in
an infinite tower of terms, and the ellipsis at the end of the equation
represents a similar pseudoscalar term plus kinetic terms.
In the large $N_c$ limit for $\mu\ll \Lambda$ it was argued that
the generalised NJL model in equation \ref{gen} is equivalent to the minimal
standard model in the sense that the two models occupy the same
parameter space. In other words $m_t$ and $m_H$ are arbitrary.

The above analysis seems to pour cold water over the whole top
quark condensate approach. If such models are merely re-statements
of the standard model (or supersymmetric standard model) then
why bother? However all is not lost since if one can reduce the scale
of new physics down to the TeV region then it {\em will}
be possible to distinguish the new physics which is responsible
for the top quark condensate from standard physics.
In other words if the dimensional scale associated with the
non-renormalisable operators is a TeV or so, then the whole
operator approach will begin to break down, just as
Fermi theory of weak interactions breaks down at energies
of order the $W$ mass. In this case the Hasenfratz {\it et al}
operator approach is inadequate, and one must confront the new
physics responsible for the non-renormalisable operators
directly. Of course new physics at a TeV is dangerous because
we know that it can potentially lead to FCNC's.
However there are models in the literature
based on heavy gauge boson exchange (of mass $\sim $ 1 TeV)
which can lead to a top quark condensate without
excessive FCNC's due to a GIM mechanism (Glashow, Illiopoulos and
Maiani 1970). These models are discussed in
King 1992a, 1992b, Elliott and King 1992, 1993
and Evans et al 1993a.
Other models without a GIM mechanism include for example
Hill (1991), King (1991), Clague and Ross (1991),
Lindner and Lust (1991), Lindner and Ross (1992).

\section{Conclusion}

In this review we have traced the development of the
subject of dynamical symmetry breaking from the
invention of technicolour in 1979 through the idea
of top quark condensates a decade later
and up to the present day. We have discussed the
simple idea of technicolour, which breaks electroweak
symmetry by analogy with QCD, but which does not
provide fermion masses. This led us to consider extended
technicolour and its phenomenological problems.
However since the time that the review
by Farhi and Susskind (1981) was written
there has been real progress in tackling the problems
of extended technicolour,
and it has been one of the purposes of the present review
to discuss this progress. Much of the progress has been concerned
with the idea of condensate enhancement, which enables
fermion and pseudo-Goldstone boson masses to be enhanced,
thereby allowing the extended technicolour breaking scale to be
raised and flavour-changing neutral currents to be suppressed.
Condensate enhancement may be achieved in either walking
technicolour theories, or strong extended technicolour theories,
and both these ideas have been reviewed in as non-technical
a way as possible. The basic message of this approach is that
it is possible to enhance the condensate sufficiently to
avoid problems with flavour-changing neutral currents,
but the price that must be paid appears to be fine-tuning
of one sort or another. Walking technicolour is not a generic
property of technicolour models, and so some tuning in the space
of models is required. In strong extended technicolour theories,
more direct fine-tuning is required and such fine-tuning is
accompanied by light scalar bound states.

We have also discussed the experimental prospects for technicolour
theories. We have seen that the sure-fire way of testing technicolour
theories is to perform longitudinal gauge boson scattering
experiments at TeV energies. This is analogous to doing pion-pion
scattering experiments and as in that case, we would expect
a resonance structure of some sort which will hold the key
to the strong dynamics responsible for
electroweak symmetry breaking.
A more indirect indication of technicolour would be the
discovery of pseudo-Goldstone bosons, which we have also
discussed. A third way to look for evidence of technicolour
is in precision electroweak measurements which are
sensitive to new physics. We have seen that current data
disfavour models with too many new heavy fermions,
such as the single techni-family, but it is impossible
to rule out the technifamily this way because of calculational
uncertainties, and unconventional dynamics which can lead
to effects which partly cancel the large corrections.
Finally, low-scale technicolour is a fairly recent idea which we
have chosen to highlight because it can lead to spectacular
technicolour signatures at LEP and the Tevatron.

We have also discussed the recent idea of top quark condensates.
This idea is initially quite an attractive one since it
promises to account for the large top quark mass without
relying on extended technicolour. However it turns out that
the top quark mass always comes out too large in the simplest
models, even for a scale of new physics equal to the Planck scale,
and in addition there appears to be a severe fine-tuning problem
associated with keeping the top quark mass much lower
than the natural scale of the operators which must be introduced.
It was in fact claimed that such models are in fact
isomorphic to the standard model, a claim which seems
to pull the rug from under this approach.
One solution to all these problems is to bring down the scale
of the new physics responsible for the top quark condensate
down to a TeV or so. Of course this means that the top quark condensate
cannot break electroweak symmetry by itself otherwise the top
quark would come out to be much too heavy;
electroweak symmetry must be broken by some
other physics in addition to top quark condensates.
One example is the idea of fourth family condensates.
Given all these ideas it is important to try and put them into practice
and construct explicit models of dynamical electroweak symmetry breaking.
There are so many models in the literature that it would
have been impossible to review them all,
and since there is no leading candidate dynamical model
we have focussed on only one or two
models simply because the author is familiar with them.

At the time of writing in 1994 the standard model
looks in remarkably good shape. The theory agrees
with experiment at the 1\% level, and the
evidence for the discovery of the top
quark has just been announced. However
the Higgs boson of the
standard model has not been discovered,
and the mechanism of electroweak symmetry breaking
remains untested experimentally.
This lack of experimental information about the mechanism
of electroweak symmetry breaking is accompanied by a
general feeling of theoretical
dissatisfaction with the single Higgs doublet of the
minimal standard model.
The hope is that there is some new physics beyond the standard model
which plays an important part in electroweak symmetry breaking.
The question is what is the nature of this new physics?

Although in this article we have only discussed the possibility
that the effect of the Higgs doublet is replaced
by some kind of fermion-antifermion condensate, it is worth
emphasising that there are other alternatives. We have already mentioned
that the naturalness problems accompanying elementary Higgs scalars
may be eliminated by postulating an $N=1$ supersymmetry
which is broken at the TeV scale.
It is also perhaps worth mentioning
the ideas of composite quarks and leptons, and composite
gauge bosons. Alternatively the new physics may involve some
combination of the above ideas, or possibly some new ingredient
which we have not yet imagined.

The problem of electroweak symmetry breaking,
or the problem of the origin of mass,
is the number one problem of late twentieth century physics.
 From our present viewpoint in 1994 the answer to this question
remains as open today as it always was, but
it is a question whose answer lies at the TeV energy scale.
It will therefore take a supercollider such as the
planned large hadron collider (LHC),
which would occupy the LEP tunnel at CERN,
to answer this question.

\vspace{0.5in}

{\bf Acknowledgements}

My thanks to my colleague Douglas Ross
and former students Helen Chestermann,
Terry Elliott, Nick Evans and Samjid Mannan
who have collaborated on much of my work
described here. A special thanks to
Douglas Ross and Peter White
for reading the manuscript and Nick Evans for
keeping me up to date on the subject of
precision electroweak measurements.
Finally I would like to thank all those
who have contributed to dynamical electroweak
symmetry breaking and whose work is
not directly acknowledged below.

\newpage

{\bf \Large References}

\noindent

Abe F {\it et al} ( CDF collaboration )1994
{\it FERMILAB-PUB-94/097-E}

Altarelli et al 1992 {\it Nucl. Phys. }{\bf B 369} 3

Altarelli et al 1993 {\it Nucl. Phys. }{\bf B 405} 3

Appelquist T , Karabali D and Wijewardhana L 1986
{\it Phys. Rev. Lett.} {\bf 57} 957

Appelquist T and Wijewardhana L 1987
{\it Phys. Rev.} {\bf D35} 774

Appelquist T , Carrier D, Wijewardhana L and
Zheng W  1988a {\it Phys. Rev. Lett. }{\bf 60} 1114

Appelquist T, Lane K and Mahanta U 1988b
{\it Phys. Rev. Lett.} {\bf 61} 1553

Appelquist T, Soldate M, Takeuchi T and Wijewardhana L 1988c
in {\it Tev Physics}, proc. 12th Johns Hopkins Workshop, Baltimore,
Maryland (ed. G. Domokos and S. Kovesi-Domokos, pub. World
Scientific,1989)

Appelquist T, Takeuchi T, Einhorn M
and Wijewardhana L 1989 {\it Phys. Lett. }{\bf B220} 223

Appelquist T and Shapira O 1990
{\it Phys. Lett.}{\bf B249} 83

Appelquist T , Terning J and Wijewardhana L 1991
{\it Phys. Rev.}{\bf D44}

Appelquist T and Triantaphyllou G 1992a
{\it Phys. Lett. }{\bf B278} 345

Appelquist T and Triantaphyllou G 1992b
YCTP-P26-92

Appelquist T and Terning J 1994
YCTP-P21-93, BUHEP-93-23.

Babu K and Ma E 1988
{\it Phys. Rev. Lett. }{\bf 61} 674

Babu K and Ma E and Pantaleone 1989
{\it Phys. Lett. }{\bf B218} 233

Bagger J, Dawson S and Valencia G 1991
{\it Phys. Rev. Lett. }{\bf 67} 2256

Bardeen J, Cooper L and Schrieffer J 1962
{\it Phys. Rev.}{\bf 106} 162

Bardeen W, Leung C and Love S 1986
{\it Phys. Rev. Lett.}{\bf 56} 1230

Bardeen W, Hill C and Lindner M 1990a
{\it Phys. Rev.} {\bf D41} 1647

Bardeen W , Love S and Miransky V 1990b
{\it Phys. Rev.}{\bf D42} 3514

Burgess C, Godfrey S, Konig H, London D and Maksymyk I 1993
``A Global Fit to Extended Oblique Parameters''
{\it McGill-93/24}

Cahn R 1989
{\it Rep. Prog. Phys.} {\bf 52} 389

Cahn R and Suzuki M 1991
{\it Phys. Rev.}{\bf D44} 3641

Carena M, Clark T, Wagner C, Bardeen W and Sasaki 1992
{\it Nucl. Phys.}{\bf B369} 33

Casalbuoni R, DeCurtis S, Dominici D and Gatto R 1987
{\it Nucl. Phys. }{\bf B282} 235

Casalbuoni R, DeCurtis S, Dominici D, Feruglio F and Gatto R 1991
{\it Phys. Lett. }{\bf B269} 361

Chanowitz M and Gaillard M 1985
{\it Nucl. Phys.} {\bf B261} 379

Chesterman H , King S and Ross D 1991
{\it Nucl. Phys.}{\bf B358} 59

Chesterman H and King S 1992
{\it Phys. Rev. }{\bf D45} 297

Chivukula R , Georgi H and Randall L 1987
{\it Nucl. Phys. }{\bf B292} 93

Chivukula R , Cohen A and Lane K 1990
{\it Nucl. Phys.}{\bf B343} 554

Chivukula R, Dugan M and Golden M 1992a
{\it Phys. Lett. }{\bf B292 } 435

Chivukula R, Selipsky S and Simmons E 1992b,
{\it Phys. Rev. Lett.}{\bf 69} 575

Chivukula R, Simmons E and Terning J 1994,
{\it hep-ph/9404209}

Clague D and Ross G 1991
{\it Nucl. Phys. }{\bf B364} 43

Clark T, Love S and Bardeen W 1990
{\it Phys. Lett. }{\bf B237} 235

Curtis D and Pennington M 1993
{\it Phys. Rev.} {\bf D48} 4933

Dagotto E , Hands S ,  Kocic A and Kogut J 1990
{\it Nucl. Phys. }{\bf B347} 217

Dawson S and Willenbrock S 1989
{\it Phys. Rev. }{\bf D40} 2880

Dimopoulos S and Susskind L 1979 {\it Nucl. Phys.} {\bf B155} 237

Dimopoulos S 1980 {\it Nucl. Phys.}{\bf B168} 69

Dimopoulos S, Raby S and Susskind L 1980
{\it Nucl. Phys. } {\bf B169} 373

Dimopoulos S and Raby S 1981
{\it Nucl. Phys.}{\bf B192} 353

Dobado A, Espriu D and Herrero M 1991
{\it Phys. Lett. }{\bf B255} 405

Eichten E and Lane K 1980 {\it Phys. Lett.} {\bf B90} 125

Eichten E, Hinchliffe I, Lane K and Quigg C 1986,
{\it Phys. Rev} {\bf D34} 1547

Eichten E and Lane K 1989 {\it Phys. Lett.} {\bf B222} 274

Elliott T and King S 1992
{\it Phys. Lett. }{\bf B283} 371

Elliott T and King S 1993
{\it Z Phys. }{\bf C58} 609

Ellis J, Gaillard M, Nanopoulos D and Sikivie P 1981,
{\it Nucl. Phys.}{\bf B182} 529

Ellis J , Fogli G and Lisi E 1992
{\it Phys. Lett.} {\bf B292} 427

Evans N, King S and Ross D 1993a
{\it Z Phys.} {\bf C60} 509

Evans N , King S and Ross D 1993b
{\it Phys. Lett.} {\bf B303} 295

Evans N and Ross D 1993c
SHEP-92/93-23 {\it Nucl. Phys.}{\bf B} (to appear)

Evans N 1993
``V,W and X in Technicolour Models'' {\it SWAT/16}

Evans N 1994
SWAT/27, {\it Phys. Lett.}{\bf B} (to appear)

Farhi E and Susskind L 1981 {\it Phys. Rep.} {\bf 74} 277

Farhi E and Jackiw R 1982
{\em Dynamical Gauge Symmetry Breaking: a collection of reprints}
(published by World Scientific)

Foot R and Lew H 1990
{\it Phys. Rev. }{\bf D41} 3502

Gates E and Terning J 1991
{\it Phys. Rev. Lett.}{\bf 67} 1840

Georgi H 1984 {\em Weak Interactions and Modern Particle Theory}
(published by Benjamin-Cummings)

Georgi H 1990 {\em ``Technicolour and Families''},
Proc. 1990 Int. Workshop on Strong Coupling Gauge Theories and Beyond,
July 28-31, 1990, Nagoya, Japan (ed. T. Muta and K. Yamawaki,
pub. World Scientific 1991)

Ginzburg V and Landau L 1950
{\it Zh. Eksp. Teor. Fiz.}{\bf 20} 1064

Giudice G and Raby S 1992
{\it Nucl. Phys. }{\bf B}

Glashow S , Iliopoulos I and  Maiani L 1970
{\it Phys. Rev.} {\bf D2} 1285

Golden M and Randall L 1990
{\it Nucl. Phys.}{\bf B361} 3

Hands S, Kocic A and Kogut J 1991
{\it Phys. Lett.}{\bf B273} 111

Hasenfratz A {\it et al} 1991 {\it Nucl. Phys.}{\bf B365} 79

Hill C 1981
{\it Phys. Rev.}{\bf D24} 691;
{\it ibid} 1991
{\it``Dynamical Symmetry Breaking of the Electroweak Interactions
and the Renormalisation Group''},
Proc. 1990 Int. Workshop on Strong Coupling Gauge Theories and Beyond,
July 28-31, 1990, Nagoya, Japan (ed. T. Muta and K. Yamawaki,
pub. World Scientific 1991)

Hill C and Paschos E 1990
{\it Phys. Lett.}{\bf B241} 96

Hill C , Luty M and Paschos E 1991
{\it Phys. Rev. }{\bf D43} 3011

Hill C 1991
{\it Phys. Lett. }{B266} 419

Holdom B 1981 {\it Phys. Rev.} {\bf D24} 1441

Holdom B 1985
{\it Phys. Lett. } {\bf B150} 301

Holdom B 1988a
{\it Phys. Lett. }{\bf B213} 365

Holdom B 1988b, {\it Phys. Rev. Lett.} {\bf 60} 1233

Holdom B 1989 {\it Phys. Lett. }{\bf B226} 137

Holdom B , Terning J and Verbeek K 1989
{\it Phys. Lett. }{\bf B232} 351

Holdom B and Terning J 1990
{\it Phys. Lett. }{\bf B247} 88

Kagan A and Samuel S 1991
{\it Phys. Lett. }{\bf B270} 37

Kamli A and Ross D 1992
{\it Nucl. Phys.} {\bf B373} 169

Kennedy D and Lynn B 1989
{\it Nucl. Phys. }{\bf B322} 1

King S 1989a {\it Nucl. Phys.} {\bf B320} 487

King S 1989b {\it Phys. Lett.} {\bf B229} 253

King S and Ross D 1989 {\it Phys. Lett.}{\bf B228} 363

King S and Ross D 1990 {\it Phys. Lett.}{\bf B236} 327

King S and Mannan S 1990
{\it Phys. Lett.}{\bf B241} 249

King S 1991
{\it Phys. Lett. }{\bf B263} 97

King S and Mannan S 1991
{\it J. Mod. Phys.}{\bf A6} 4949

King S and Mannan S 1992
{\it Nucl. Phys. }{\bf B369} 119

King S 1992a
{\it Phys. Rev.}{\bf D45} 990

King S 1992b
{\it Phys. Rev.}{\bf D46} 4097

King S 1992c
{\it Phys. Lett. }{\bf B281} 295

King S 1992d
{\it Phys. Rev.} {\bf D46 } R4804

King S 1993
{\it Phys. Lett. }{\bf B314} 364

Kondo K, Mino H and Yamawaki K 1989
{\it Phys. Rev. }{\bf D39} 2430

L3 Collaboration 1993 {\it Phys. Lett. }{\bf B295}  337

Lane K 1974 {\it Phys. Rev. }{\bf D10} 2605

Lane K 1994 Talk at the ICHEP 94, Glasgow.

Lindner M and Lust D 1991
{\it Phys. Lett.}{\bf B272} 91

Lindner M and Ross D 1992
{\it Nucl. Phys. }{\bf B370} 30

Lubicz V 1993
{\it Rome ``La Sapienza'' preprint 925}

Luty M 1990
{\it Phys. Rev. }{\bf D41} 2893

Luty M 1992 {\it Phys. Lett. }{\bf B292} 113

Luty M and Sundrum R 1993
{\it Phys. Rev. Lett.}{\bf 70} 529

Marciano W 1989
{\it Phys. Rev. Lett.}{\bf 62} 2793

Miransky V, Tanabashi M and Yamawaki Y 1989
{\it Phys. Lett. } {\bf B221} 177;
{\it ibid} 1989 {\it Mod. Phys. Lett. } {\bf A4} 1043

Miransky V 1991
{\it``Electroweak Symmetry Breaking and Dynamics of Tight Bound States''}
,Proc. 1990 Int. Workshop on Strong Coupling Gauge Theories and Beyond,
July 28-31, 1990, Nagoya, Japan (ed. T. Muta and K. Yamawaki,
pub. World Scientific 1991)

Mohapatra R and Zhang X 1993
{\it Phys. Lett. }{\bf B305} 106

Nambu Y and Jona-Lasinio G  1961 {\it Phys. Rev.} {\bf 122} 345

Nambu Y 1989 {\it ``New Theories in Physics''}
Proc. XI Warsaw Symposium on Elementary Particle Physics
(ed. Z.Ajduk {\it et al}, pub. World Scientific 1989)

Nussinov S 1985
{\it Phys. Lett.}{\bf B165} 55;
{\it ibid} 1992
{\it Phys. Lett.}{\bf B279} 111

Pagels H and Stokar S 1979
{\it Phys. Rev. }{\bf D20} 2947

Pauss F 1990
{\it Proc. ECFA LHC Collider Workshop -- Aachten}
{\bf CERN 90-10, ECFA 90-133}

Pendleton B and Ross G 1981
{\it Phys. Lett. }{\bf B98} 291

Peskin M 1980
{\it Nucl. Phys. }{\bf B175} 197

Peskin M 1982 Les Houches Lectures
(ed.J. B. Zuber and R. Stora, pub. North-Holland 1984)

Peskin M and Renken R 1983
{\it Nucl. Phys.} {\bf B211} 93

Peskin M and Takeuchi T 1990
{\it Phys. Rev. Lett.} {\bf 65} 964;
{it ibid} 1992 {\it Phys. Rev.} {\bf D46} 381

Politzer H 1976 {\it Nucl. Phys. }{\bf B117} 397

Preskill J 1981 {\it Nucl. Phys. }{\bf B177} 21

Randall L 1993 {\it Nucl. Phys. }{\bf B403} 122

Samuel S 1990
{\it Nucl. Phys.} {\bf B347} 625

Simmons E 1989
{\it Nucl. Phys. } {\bf B312} 253

Sundrum R and Hsu S 1993
{\it Nucl. Phys. }{\bf B391} 127

Sundrum R 1993
{\it Nucl. Phys. }{\bf B395} 60

Susskind L 1979 {\it Phys. Rev.} {\bf D20} 2619

Suzuki M 1990a
{\it Phys. Rev.} {\bf D41} 3457

Suzuki M 1990b {\it Mod. Phys. Lett. }{\bf A5} 1205

Terning J 1991 {\it Phys. Rev.} {\bf D44}

Weinberg S  1979 {\it Phys. Rev.} {\bf D20} 1277

Xiao Z, Wan L, Lu G, Yang J, Xuelei W, Guo L, Yue C 1994
{\it J Phys. G } (in print)

Yamawaki Y, Bando M and Matumoto K 1986a
{\it Phys. Rev. Lett.} {\bf 56} 1335;
{\it ibid} 1986b
{\it Phys. Lett.} {\bf B178} 308

\newpage

{\Large \bf Figure Captions}

\vspace{0.25in}

{\bf Figure} 1: Energy diagram of a superconductor.

\vspace{0.25in}

{\bf Figure} 2: BCS energy gap function.

\vspace{0.25in}

{\bf Figure} 3: Corrections to the $W^{\pm}$ propagator
from Goldstone boson exchange.

\vspace{0.25in}

{\bf Figure} 4: Mixing between the $W^0$ and the $B$ gauge bosons.

\vspace{0.25in}

{\bf Figure} 5: Extended technicolour gauge boson vertices which
can connect fermions (f) and technifermions (T) in all possible ways.

\vspace{0.25in}

{\bf Figure} 6: Radiative mechanism for extended technicolour generation
of fermion mass. The exchanged gauge boson is a heavy
ETC boson of mass $M_{ETC}$.

\vspace{0.25in}

{\bf Figure} 7: Dangerous neutral flavour-changing diagram which can
contribute to $K^0-\bar{K^0}$ mixing.
The exchanged gauge boson is a heavy
ETC boson of mass $M_{ETC}$.

\vspace{0.25in}

{\bf Figure} 8: Diagrams responsible for generating the quark and charged
lepton masses in this simple extended technicolour model.

\vspace{0.25in}

{\bf Figure} 9: The diagrammatic representation of the gap equation.

\vspace{0.25in}

{\bf Figure} 10: The numerical solutions of the gap equation for
(a) a normal QCD-like theory, (b) a Walking TC theory,
(c) a Fixed Point theory.
Both $\Sigma (p^2)$ and ${\alpha}_{TC}(p^2)$ are plotted.
The solutions can be simply described.
For $p<{\Lambda}_{TC}$, $\Sigma(p)=\Sigma(0)$ (flat).
For $p>\mu$ , where perturbation theory is more trustworthy,
the behaviour of $\Sigma(p)$ depends upon the value of
$\alpha_{TC}(p)$. In the region where
$\alpha_{TC}(p)\approx \alpha_{TC}^c$ ,
$\Sigma(p)\sim 1/p$ (see for example Peskin 1982).
However for
$\alpha_{TC}(p)\ll \alpha_{TC}^c$ ,
$\Sigma(p)\sim 1/p^2$ (up to logarithms)
as shown by Lane (1974) and Politzer (1976).
In the intermediate region denoted as a fuzzy band
the fall-off of $\Sigma$ is intermediate between these two cases.
In the case of normally running theories (figure 10a) the
logarithmic fall-off of $\alpha_{TC}$ is sufficiently fast
to ensure that the asymptotic solution sets in almost
immediately, leading to a value of the condendsate
corresponding  to $\gamma \approx 0$. In walking TC
theories (figure 10b) the coupling falls off more slowly
than a logarithm, and hence the solution $\Sigma(p) \sim 1/p$
persists over a larger range of $p$, resulting in an enhanced
condensate with $0< \gamma <1$ (Appelquist {\it et al} 1986, 1987).
In the case of a
fixed point (figure 10c) $\alpha_{TC}(p)=\alpha_{TC}^c$
over the whole range of $p>\mu$, and maximum condensate
enhancement is achieved with $\gamma \approx 1$.

\vspace{0.25in}

{\bf Figure} 11: Additional contributions to the right-hand side
of the gap equation in figure 9 coming from
(a) higher order corrections, (b) QCD gluon exchange,
(c) heavy ETC gauge boson exchange, (d) a feed-back mechanism
involving an internal top quark propagator.

\vspace{0.25in}

{\bf Figure} 12: Criticality curve for quenched QED with a four-fermion
operator.

\vspace{0.25in}

{\bf Figure} 13: Signal and background events
expected at the LHC from the decay of a
1.5 TeV technirho into $WZ$ in (a) the BESS model, and
(b) the DHT model. In both cases the signal
plus backgound (upper histogram) is clearly
visible above the background.

\vspace{0.25in}

{\bf Figure} 14: Spectrum of pseudo-Goldsone bosons
(sometimes called technipions) in the one-family
$SU(N)_{TC}$ model.

\vspace{0.25in}

{\bf Figure} 15: Spectrum of pseudo-Goldsone bosons
(sometimes called technipions) in the one-family
$SO(N)_{TC}$ model.

\vspace{0.25in}

{\bf Figure} 16: A contribution to the $Zb\bar{b}$ vertex
from neutral ETC gauge boson exchange.

\vspace{0.25in}

{\bf Figure} 17: The gap equation in the top quark condensate model.

\end{document}